\documentclass[english]{article}
\usepackage[T1]{fontenc}
\usepackage[latin9]{inputenc}
\usepackage{geometry}
\usepackage{babel}
\usepackage{prettyref}
\usepackage{float}
\usepackage{amsmath}
\usepackage{amsthm}
\usepackage{amssymb}
\usepackage{graphicx}
\usepackage[unicode=true]
 {hyperref}

\makeatletter

\providecommand{\tabularnewline}{\\}

\numberwithin{equation}{section}
\numberwithin{figure}{section}

\pdfoutput=1

\@ifundefined{showcaptionsetup}{}{%
 \PassOptionsToPackage{caption=false}{subfig}}
\usepackage{subfig}
\makeatother

\begin{document}

\title{Dynamic Principal Component Analysis: Identifying the Relationship
between Multiple Air Pollutants.}

\author{Oleg Melnikov$^{\dagger,a}$, Loren H. Raun$^{a,b}$, Katherine B.
Ensor$^{a}$\thanks{$^{a}$Rice University, Department of Statistics, Houston, TX, USA\protect \\
$^{b}$City of Houston Health and Human Services Bureau of Pollution
Control and Prevention, Houston, TX, USA\protect \\
$\dagger$Corresponding author. Tel.: +1 323 205 6534, E-mail address:
xisreal@gmail.com, http://Oleg.Rice.edu }}
\maketitle
\begin{abstract}
The dynamic nature of air quality chemistry and transport makes it
difficult to identify the mixture of air pollutants for a region.
In this study of air quality in the Houston metropolitan area we apply
dynamic principal component analysis (DPCA) to a normalized multivariate
time series of daily concentration measurements of five pollutants
(O3, CO, NO2, SO2, PM2.5) from January 1, 2009 through December 31,
2011 for each of the 24 hours in a day. The resulting dynamic components
are examined by hour across days for the 3 year period. Diurnal and
seasonal patterns are revealed underlining times when DPCA performs
best and two principal components (PCs) explain most variability in
the multivariate series. DPCA is shown to be superior to static principal
component analysis (PCA) in discovery of linear relations among transformed
pollutant measurements. DPCA captures the time-dependent correlation
structure of the underlying pollutants recorded at up to 34 monitoring
sites in the region. In winter mornings the first principal component
(PC1) (mainly CO and NO2) explains up to 70\% of variability. Augmenting
with the second principal component (PC2) (mainly driven by SO2) the
explained variability rises to 90\%. In the afternoon, O3 gains prominence
in the second principal component. The seasonal profile of PCs' contribution
to variance loses its distinction in the afternoon, yet cumulatively
PC1 and PC2 still explain up to 65\% of variability in ambient air
data. DPCA provides a strategy for identifying the changing air quality
profile for the region studied.
\end{abstract}

\section{\label{sec:Intro}Introduction}

Chemical processes are complex and nonlinear. Their dependency structures
are contaminated with cross and auto correlations, seasonality, diurnal
cycles, outliers, and noise. Direct data visualization or even basic
statistical summaries are unable to reveal the key underlying patterns
and distributions of the mixtures of air pollutants. Multivariate
data analysis (MDA) has been effectively utilized in discovering these
latent structures. Principal component analysis (PCA) is one such
tool that can identify linearly related variables that describe most
of the variability in the data. 

Recently PCA has gained traction in the study of air quality (AQ).
Buhr \cite{buhr_evaluation_1995} used PCA to examine sources of nitrogen
oxides ($\text{NO}_{x}$) and carbon monoxide (CO) from other pollutants.
Trainer \cite{trainer_review_2000} studied formation and loss of
ozone ($\text{O}_{3}$) through PCA and bivariate regression of pollutants.
Gonçalves \cite{goncalves_effects_2005} related child morbidity and
meteorology patterns to ambient AQ through PCA of pollutants and meteorological
factors.

Many authors recognize the seasonal characteristics of environmental
data and analyze winter and summer observations separately. Pissimanis
\cite{pissimanis_spatial_2000} applied PCA to examine spatial distribution
of $\max\left(\text{O}_{3}\right)$ concentrations in the summer months.
Álvarez \cite{alvarez_spatial_2000} applied rotated PCA to assess
spatio-temporal variability in winter and summer. Statheropoulos \cite{m._statheropoulos_principal_1998}
related key principal components (PC) to emissions and ozone  via
PCA on winter and summer data. 

Some other authors recognize the diurnal pattern of the air pollution
data. Buhr \cite{buhr_assessment_1992} contrasted air pollution to
emission ratios with the help of PCA performed on morning data. Abdul-Wahab
\cite{abdul-wahab_principal_2005} employed PCA to construct uncorrelated
components based on air pollution and environmental data separately
aggregated for the day and night hours. Lengyel \cite{lengyel_prediction_2004}
examined day and night AQ via PCA of air pollution and meteorological
observations. Sousa \cite{sousa_multiple_2007} exploited hourly air
pollution data and meteorology to construct the components.

Still most analyses ignore the non-stationary structure of environment
AQ data \cite{yu_time_2015}. Since PCA assumes fixed distribution
parameters, an application of static PCA on observations from a distribution
with time-dependent parameters is not appropriate. While dynamic PCA
variants have been applied to chemical processes (\cite{ku_disturbance_1995}),
climatology (\cite{kim_comparison_1999}), (to our knowledge) it has
not been used to study air pollution until now. 

We construct DPCA components on a two dimensional time domain (hours
of a day $\times$ days of studied time period) and investigate the
organization of principal components and their contribution to overall
variability. We define DPCA as a moving window static PCA. Such form
of DPCA was studied by \cite{jeng_adaptive_2010}, \cite{liu_moving_2009},
\cite{wang_process_2005} and applied to electroencephalography in
\cite{xie_dynamic_2014}.

The novelty of this paper is its application of DPCA to air pollution
observations with the objective to
\begin{enumerate}
\item Demonstrate a proper application of PCA technique to cyclostationary
time-series
\item Approximate non-linear dependence with a linear technique
\item Assess absolute and relative performance of such application
\item Interpret linearity between PCA input variables and translate it to
original AQ indicators
\item Reveal diurnal and seasonal patterns of strong and weak linear dependence
among PCA input variables
\end{enumerate}
This paper stops short of use of the identified dynamic PCs in forecasting,
construction of air quality indicators (AQI), dimension reduction,
etc. Some of the aforementioned papers (and references therein) have
already demonstrated such extensions to PCA. Also, we do not account
for spatial information, which has been investigated by other authors
(e.g. \cite{alvarez_spatial_2000,pissimanis_spatial_2000}). Instead,
we construct spatially-averaged observations (SAO) to achieve a greater
degree of robustness.

The paper is organized as follows. \prettyref{sec:PCA} discusses
PCA assumptions, methodology and interpretation. It also defines DPCA
and a notation helpful when referencing dynamic factors. \prettyref{sec:Data}
describes data pre-treatment, determines a suitable transformation,
and verifies dynamic correlations to justify the use of DPCA. In \prettyref{sec:Results-and-discussion}
we apply DPCA to construct an informative 3D profile, identifying
the contribution to the explained variability of each PC. We then
evaluate contributions averaged across each hour and study dynamic
PC loadings for two times of a day, namely 7am and 2am. Finally, we
compare our DPCA efforts to employment of static PCA on air pollution
data. We close with a short section of concluding remarks.    

\section{\label{sec:PCA}Methodology}

\subsection{Assumptions}

PCA assumes the distribution of a data matrix $X$ is characterized
by constant mean and covariance parameters. In other words, since
PCs are linear combinations of input variables (columns of $X$),
the latter must be linearly related on the full observational interval
\cite{shlens_tutorial_2014}. This condition is problematic, since
most observed processes are not linearly related and their distribution
parameters may change with state, space, or time (even if the distribution
family remains the same). For example, environmental and meteorological
data often exhibit trend non-stationarity as the process mean exhibits
seasonal and diurnal patterns. Fortunately, this behavior, termed
cyclostationarity, still exhibits stationarity on a neighborhood of
any point of a cycle. This local stationarity can be tested and local
observations can be further explored with the usual PCA \cite[p.314]{jolliffe_principal_2002},
\cite{kim_comparison_1999,davis_outlier_2006}. Similarly, in this
paper, we perform static PCA on a fixed-size window, sliding in time
along observations. This yields time-changing (dynamic) PCs on samples
that are sufficiently small to remain weakly stationary, but still
seize the local dependence structure\@.

Still, there is a body of literature discussing the assumption of
whether $X$ must have independent and identically distributed (iid)
rows, each of which are multivariate normal (MVN) for PCA to make
sense \cite[p.19]{jolliffe_principal_2002}, \cite[p.229]{mardia_multivariate_1979},
\cite[p.488]{anderson_introduction_2003}, \cite[p.102]{jackson_users_2005}.
The authors determine that theoretical derivations, descriptive use
of PCA, and most results of a sample PCA do not require normality.
In the case of time series, a weak stationarity of $X$ is usually
sufficient for consideration of the consistent estimates of the first
two moments of the distribution of $X$ \cite[p.485]{ruey_s._tsay_analysis_2010},
\cite[p.365]{jackson_users_2005}. The assumption of normality adds
an additional meaning to  the inferred PCs. An interested reader
may not that in some disagreement, a few authors imply that MVN assumption
is important \cite[p.558]{friedman_elements_2001}, \cite[p.151]{korkmaz_mvn:_2014},
some claim that MVN assumption can be omitted altogether \cite[p.39]{jolliffe_principal_2002},
\cite[p.490]{ruey_s._tsay_analysis_2010}, some develop alternative
approximations to overcome the MVN assumption \cite{qian_principal_1994},
and most simply proceed with PCA without explicitly noting any assumptions.
We use and test normality only to determine the robustness point at
which data outliers become insignificant.

\subsection{\label{subsec:Robustness}Robustness}

PCA, as a least squares method, is dangerously sensitive to outliers.
These ``atypical'' observations may significantly affect estimation
of the components of the analysis, such as the eigenvectors and eigenvalues
of the covariance matrix of $X$. PCA robustness can be achieved in
a variety of ways, ranging from the least-recommended removal of peripheral
observations (or even variables) and transformation of the input data
to robustifying the intermediate covariance matrix or the terminal
PC components \cite[p.233]{jolliffe_principal_2002}, \cite[p.365]{jackson_users_2005},
\cite{wold_principal_1987}. 

In practice, real world environmental data often exhibits ill-suited
skewness and can be ``symmetricized'' with several favored non-linear
transformations, such as logarithms, roots, powers (e.g. Box-Cox transform),
ratios, log differences, reciprocals, logit transforms (of proportions),
and alike \cite{georgopoulos_statistical_1982,piegorsch_analyzing_2005}.
This data pre-treatment often coincides with \emph{normalization}
(herein defined as aligning data to MVN), which can, in turn, be checked
with a battery of statistical tests. Among popular MVN tests are those
developed by Mardia, Henze-Zirkler's and Royston. Mardia's skewness
and kurtosis tests give a greater insight on the shape fit to MVN
\cite{hair_multivariate_2006,korkmaz_mvn:_2014}.  We use one such
MVN test to identify a suitable transformation for our data. With
air pollution data, in particular, natural logarithm of some or all
variables helps stabilize asymmetric variability and diminish the
effect of extreme events \cite{georgopoulos_statistical_1982,abdul-wahab_principal_2005,buhr_assessment_1992,davis_outlier_2006,parrish_carbon_1991}.

\subsection{\label{subsec:Definition-and-interpretation}Definition and interpretation}

Consider a centered data matrix $X=\left[x_{np}\right]\in\mathbb{R}^{\mathfrak{n}\times\mathfrak{p}}$
with $\mathfrak{n}$ observations and $\mathfrak{p}$ variables, where
each row follows the same multivariate, but not necessarily normal,
distribution with fixed mean and variance parameters, estimated as
$\left(\mathbf{0},\Sigma\right)$. A (static) principal component
analysis (PCA) is defined as a linear transformation of these correlated
variables to uncorrelated principal components, $\text{PC}_{p}$,
$k=1..\mathfrak{p}$,
\begin{eqnarray}
z_{.k} & := & \text{PC}_{k}=v_{1k}x_{.1}+\cdots+v_{k\mathfrak{p}}x_{.\mathfrak{p}}\nonumber \\
 & = & \left[x_{.1}\ldots x_{.\mathfrak{p}}\right]v_{k.}=Xv_{k.}\label{eq:PCA_lin_comb}
\end{eqnarray}
where $v_{k.}'=\left[v_{1k}\ldots v_{\mathfrak{p}k}\right]\in\mathbb{R}^{1\times\mathfrak{p}}$
are the suitable loading coefficients, and $x_{.p}'=\left[x_{1p}\ldots x_{\mathfrak{n}p}\right]\in\mathbb{R}^{1\times\mathfrak{n}}$.
In other words, PCA decomposes $X$ into two component (or factor)
matrices, latent values (PC scores) and latent vectors (PC loadings).
For convenience, the components are ordered by their contribution
to the overall variability of the transformed data set. So, $\text{PC}_{1}$
has the largest contribution to variance, $\text{PC}_{2}$ - second
largest, and so on. 

One interpretation of PCA is that  in the process of decorrelation
of original variables it breaks up the entire variability of uncorrelated
PCs into summable variances represented by squared eigenvalues of
$\Sigma$. The largest eigenvalues identify principal components most
relevant to the analysis since they contain most of variability. The
smallest eigenvalues are thought to represent the noise in the data.
Hence, if the noise components are identified, a reasonable approximation
of $X$ can be recovered from the surviving dominant patterns.

Since PCA is scale-dependent, disparate units and scales of input
variables hinder interpretability of the results \cite[p.219]{mardia_multivariate_1979}.
It is, thus, common to scale raw observations in some standardized
way (usually, to mean 0 and variance 1), so that neither variable
dominates the sample covariance matrix, and, consequently, the resulting
components. Such standardization deems the input variables unitless,
thereby clouding the subsequent inference. A good rule of thumb is
to keep data in their original units, if PCA on a standardized dataset
is not significantly different from that of PCA on raw data. Also,
note that scaling up pure noise observations (with low variance) will
enhance their impact in the analysis \cite{wold_principal_1987}.

Still direct reading of PC loadings remains challenging since loading
coefficients can take negative values (weights) and void the sum-of-parts
interpretability that is prised in other popular factorization techniques,
such as negative matrix decomposition (NMF). Hence, PC loadings may
benefit from an additional transformations to ease interpretation
\cite[p.492]{ruey_s._tsay_analysis_2010}.  

\subsection{Decomposition}

The workhorse behind PCA is a singular value decomposition (SVD) of
a data matrix $X_{\mathfrak{n}\times\mathfrak{p}}$, or, equivalently,
the eigenvalue decomposition (EVD) of its sample covariance matrix,
$\Sigma_{\mathfrak{p}\times\mathfrak{p}}$. 

The former is a factorization 
\[
X_{\mathfrak{n}\times\mathfrak{p}}=U_{\mathfrak{n}\times\mathfrak{p}}\cdot\Lambda_{\mathfrak{p}\times\mathfrak{p}}\cdot V_{\mathfrak{p}\times\mathfrak{p}}'
\]
where $\Lambda$ is diagonal. $U,V$ are orthogonal , i.e. $U'U=I_{\mathfrak{p}}=V'V$
(or $U'=U^{-1}$ and $V'=V^{-1}$) . These are left and right eigenvectors
of $X$.

As with any symmetric positive semi-definite (PSD) matrix, the EVD
of $\Sigma$ is (up to a scaling factor) 
\begin{eqnarray*}
\Sigma & \propto & X'X=\left(U\Lambda V'\right)'\left(U\Lambda V'\right)=V\Lambda^{2}V'\\
\Sigma V & \propto & V\Lambda^{2}V'V=V\Lambda^{2}
\end{eqnarray*}

The components in both decompositions exist and are unique. Note that
SVD eigenvalues are equal to EVD eigenvalues. Also, right eigenvectors
of $X$ are the eigenvectors of $X'X$, and left eigenvectors of $X$
are the eigenvectors of $XX'$.

To summarize, PC transformation relates $X$ to its \emph{score} matrix
$Z$ as $Z_{\mathfrak{n}\times\mathfrak{p}}:=U\Lambda=XV$ or $z_{.k}:=u_{.k}\lambda_{k}=Xv_{.k}$,
where we index components by $k$ and variables (pollutants herein)
by $p$ ($k,p=1..\mathfrak{p}$), $v_{.k}$ are loading coefficients
from \prettyref{eq:PCA_lin_comb}, and 
\begin{itemize}
\item $\Lambda=\text{diag}\left\{ \lambda_{k}\mid0\le\lambda_{k+1}\le\lambda_{k}\right\} $
is diagonal matrix of (ordered) singular values of $\Sigma$ (in other
words,\emph{ }standard deviations of PCs). Off-diagonal zeros imply
uncorrelated PCs.
\item $\Lambda^{2}$ is a diagonal matrix of (ordered) eigenvalues of $\Sigma$
and represent the variances of PCs.
\item $V=\left[v_{.1}\ldots v_{.\mathfrak{p}}\right]=\left[v_{pk}\right]$
is a standardized PC loading matrix with columns as standardized PC
loadings of $X$, representing PC directions or eigenvectors of $\Sigma$.
The elements of $V$, $v_{pk}$, are PC loading coefficients or weights;
and, $\mathfrak{p}\times\mathfrak{p}$ matrix $V\Lambda=\left[\lambda_{1}v_{.1}\ldots\lambda_{\mathfrak{p}}v_{.\mathfrak{p}}\right]$
is $V$'s non-standardized counterpart.  
\item $U=\left[u_{.1}\ldots u_{.\mathfrak{p}}\right]_{\mathfrak{n}\times\mathfrak{p}}$
is a standardized PC score matrix with columns as\emph{ }standardized\emph{
PCs} of $X$ and rows (transformed observations) as row scores, also
termed factor scores or $z$-scores, of PCs. The $n$th element of
$z_{.k}$, $z_{nk}$, is the PC score (or factor score) of the $p$th
PC for the $n$th observation. The matrix $Z:=U\Lambda=\left[z_{.1}\ldots z_{.\mathfrak{p}}\right]$
is its non-standardized analog.
\end{itemize}
An expanded matrix notation of PCA factorization is 

\begin{equation}
\underset{X}{\underbrace{\overset{_{\text{variables}}^{\text{correlated}}}{\left[x_{.1}...x_{.\mathfrak{p}}\right]}}}\underset{V}{\underbrace{\overset{_{\text{(PC loadings)}}^{\text{eigenvectors}}}{\left[v_{.1}...v_{.\mathfrak{p}}\right]}}}=\overset{_{\text{observations}}}{\begin{bmatrix}x_{1.}'\\
\vdots\\
x_{\mathfrak{n}.}'
\end{bmatrix}}V=\underset{Z}{\underbrace{\overset{_{\text{PCs}}^{\text{uncorrelated}}}{\left[z_{.1}...z_{.\mathfrak{p}}\right]}}}=\overset{_{\text{PC scores}}}{\begin{bmatrix}z_{1.}'\\
\vdots\\
z_{\mathfrak{n}.}'
\end{bmatrix}}\label{eq:PCA_Matrix_Form}
\end{equation}
where, in a time series context, multivariate observations $x_{n.}\in\mathbb{R}^{\mathfrak{p}}$
and PC scores $z_{n.}\in\mathbb{R}^{\mathfrak{p}}$ are chronologically
indexed by time.

PCA offers some properties useful in interpretation of the results.
Explained variance (EV) is the proportion of the total variability
(of the PCs) accounted for by a specific PC. These are the diagonal
values of $\Lambda^{2}/\text{Trace}\left(\Lambda^{2}\right)$ matrix.
Variables of primary interest are EV and cumulative EV (CEV): 
\begin{eqnarray}
\text{EV}_{k} & = & \lambda_{k}^{2}/\text{Trace}\left(\Lambda^{2}\right)\label{eq:EV_def}\\
\text{CEV}_{k} & = & \sum_{i=1..k}\text{EV}_{i}\nonumber 
\end{eqnarray}
A more detailed discussion of PCA is established in \cite{mardia_multivariate_1979,friedman_elements_2001,jackson_users_2005,peres-neto_how_2005}.

\subsection{\label{subsec:DPCA_notation}DPCA notation and diagram }

Application of static PCA on a data with time-dependent structure
is unreliable, since the procedure attempts to linearly approximate
the complex non-linear relations between variables \cite{ku_disturbance_1995}.
Instead, dynamic PCA (DPCA), a simple extension of PCA, can reveal
the dynamics of the underlying data structure. Our definition of DPCA
is an application of the sample PCA on a sliding window of fixed width
$\ell$ \cite{jeng_adaptive_2010,wang_process_2005}. For a cyclostationary
time series, a local (in time) sample of observations is approximately
weakly stationary with (some) fixed distribution \cite{jolliffe_principal_2002}.
PCA applied on a windowed data captures the linear relation of the
variables. As the window slides forward at a constant rate of one
observation at a time, the time-indexed PC loadings and scores express
the overall non-linear relation. 

Since we apply PCA on a window sliding across time, all resulting
statistics are time dependent. For reasons discussed in Section \ref{subsec:SAO_section},
we consider time to be a two dimensional domain of $\text{hours}\times\text{days}$.
This avoids diurnal and seasonal non-stationarities and allows for
separate diurnal and seasonal data analysis. Whenever notation $\text{EV}_{k}$
may be ambiguous, we underline the specific time dependencies: 

\begin{eqnarray}
\text{EV}_{h.k} & := & \left[\text{EV}_{hdk}\right]_{\forall d}\in\mathbb{R}_{+}^{\mathfrak{d}}\label{eq:Dynamic_Notation}\\
\text{EV}_{..k} & := & \left[\text{EV}_{hdk}\right]_{\forall h,d}\in\mathbb{R}_{+}^{24\times\mathfrak{d}}\nonumber 
\end{eqnarray}
where $h$ is an hour of a day, $d$ is a day of the time period,
and $k$ identifies the corresponding $k$th PC. In our dataset we
have $\mathfrak{d}=\max\left\{ d\right\} $ or 1095 days. 

Similarly, dynamic PC loadings are defined via a 4 dimensional array
$V=\left[v_{hdpk}\right]\in\mathbb{R}^{24\times\mathfrak{d}\times\mathfrak{p}\times\mathfrak{p}}$
with analogous definitions $v_{h.pk}\in\mathbb{R}^{\mathfrak{d}}$,
$v_{hd.k},v_{hdp.}\in\mathbb{R}^{\mathfrak{p}}$, $v_{hd..}\in\mathbb{R}^{\mathfrak{p}^{2}}$,
$v_{.dp.}\in\mathbb{R}^{24\times\mathfrak{p}}$, $v_{h...}\in\mathbb{R}^{\mathfrak{d}\times\mathfrak{p}\times\mathfrak{p}}$,
etc.  A dot increments a dimension of the variable by the maximum
of the corresponding index placeholder. One dot designates a vector,
two - a matrix (first dot defines rows, second - columns), three -
a 3D array (third dot defines the size of the third dimension). So
$\left[v_{h.pk}\right]_{\forall p,k}$ is a $\mathfrak{p}\times\mathfrak{p}$
matrix of $\mathfrak{n}$-vectors as elements. This ameliorates visualization
of dynamic loadings and other variables. In the same way we assign
notation for dynamic PCs: $\left[\text{PC}_{hdk}\right]\in\mathbb{R}^{24\times\mathfrak{n}\times\mathfrak{p}}$,
$\text{PC}_{h.k}\in\mathbb{R}^{\mathfrak{n}}$, $\text{PC}_{..k}\in\mathbb{R}^{24\times\mathfrak{n}}$,
etc.

\begin{figure}[h]
\begin{centering}
\includegraphics[width=0.5\columnwidth]{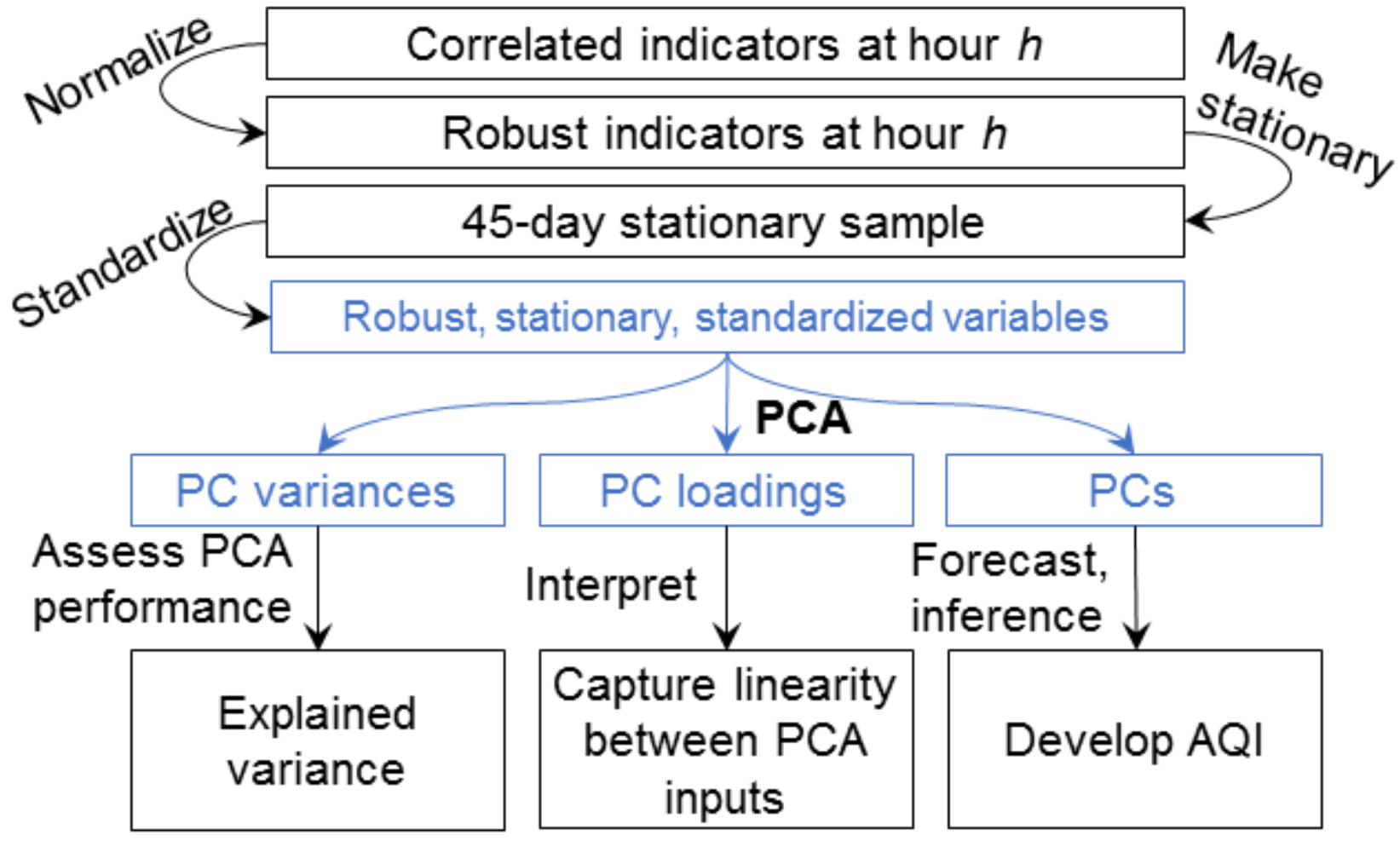}
\par\end{centering}
\caption{\label{fig:DPCA_Diagram} Application of DPCA. The correlated indicators
are spatially-averaged observations (SAO) constructed in Section \ref{subsec:SAO_section}.}
\end{figure}

Schematically, our application of DPCA is exhibited in Figure \ref{fig:DPCA_Diagram},
with an exception of forecasting.

\section{\label{sec:Data}Data}

\begin{figure}[h]
\begin{centering}
\includegraphics[width=0.5\columnwidth]{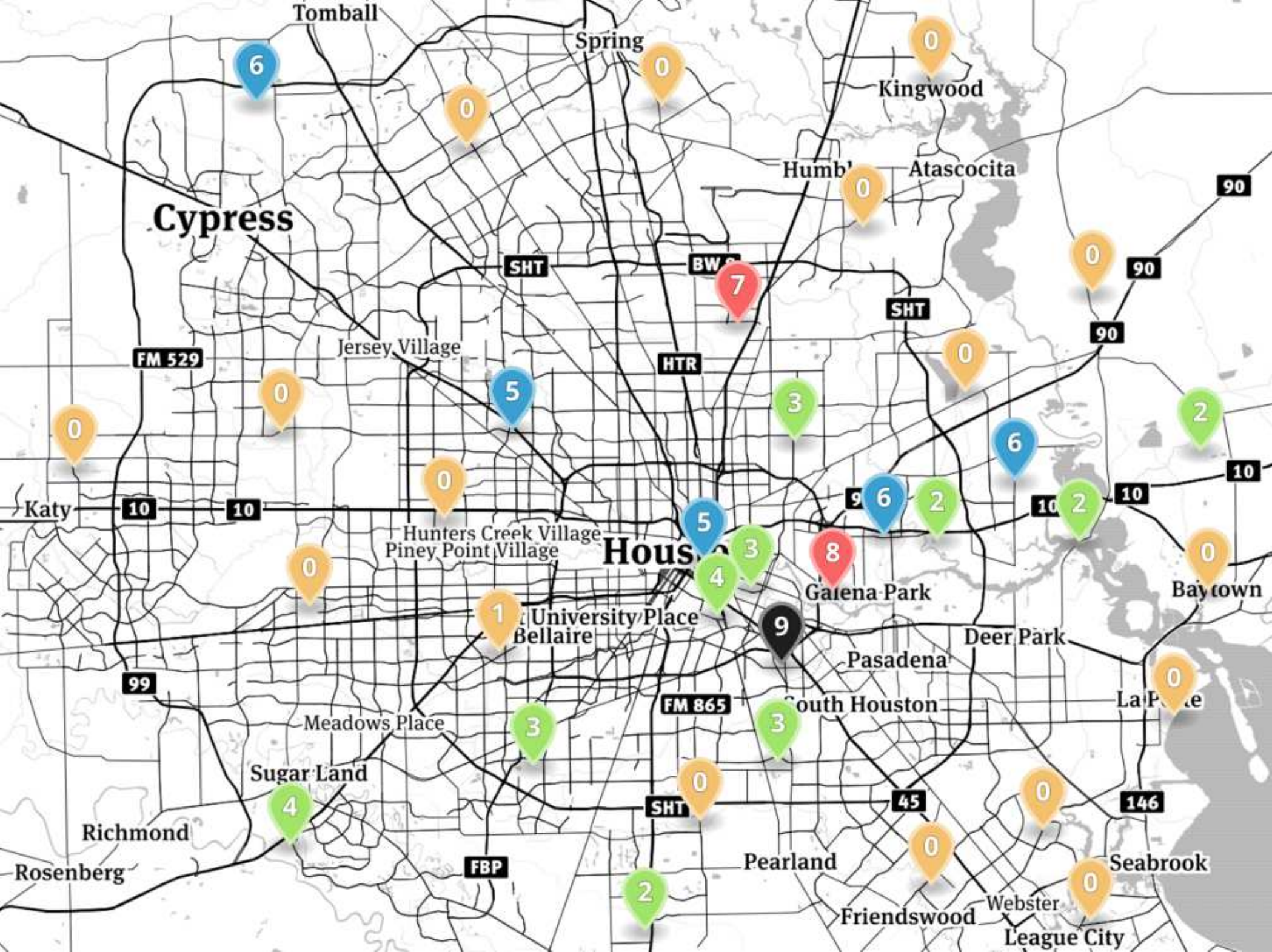}
\par\end{centering}
\caption{\label{fig:site_map} Monitoring sites in Houston, Harris County,
TX, USA. Area of study excludes some sites located near Galveston
Bay and Houston Ship Channel, or too distant from Houston. Pin colors:
black - tracks all 5 pollutants, red - 4, blue - 3, green - 2, orange
- 1. Pin numbers: 9 - tracks OCNSP (short for $\text{O}_{3}$, $\text{CO}$,
$\text{NO}_{2}$, $\text{SO}_{2}$, $\text{PM}_{2.5}$, respectively),
8 - OCNS, 7 - OCNP, 6 - ONP, 5 - OCN, 4 - OC, 3 - OS, 2 - ON, 1 -
N, 0 - O). Data source: \protect\href{http://www.tceq.state.tx.us}{www.tceq.state.tx.us}}
\end{figure}

Texas Commission on Environmental Quality (TCEQ) provides an access
to measurements of air pollutant concentrations from Texas monitoring
stations (sites).  The dataset contains hourly observations of 5
pollutants: ozone ($\text{O}_{3}$), carbon monoxide ($\text{CO}$),
nitrogen dioxide ($\text{NO}_{2}$), sulfur dioxide ($\text{SO}_{2}$),
and particulate matter less than 2.5 micrometers ($\text{PM}_{2.5}$),
from 1/1/2009 00:00 CST to 12/31/2011 23:00 CST (that is 1095 days
or 26,280 samples) collected from 35 monitoring sites throughout the
area of Houston, Texas; see Figure \ref{fig:site_map}. The study
region excludes sites that are non-representative of air pollution
profile of the Houston metropolitan area (HMA). For example, the Galveston
Bay area is an oceanic coastal line with concentrations expected to
differ from those in HMA. The Houston Ship Channel, unlike HMA, is
an industrialized home to numerous petroleum refineries, and port
and chemical manufacturing plants \cite{brioude_new_2012,kim_characterization_2005}.
Some other sites are considered too remote. Gas concentrations are
measured in (dimensionless units of) parts per billion (ppb), whereas
$\text{PM}_{2.5}$ is in $\mu\text{g}/\text{m}^{3}$. 

Spatial information is lost once we construct spatially-averaged observations
(SAO) in Section \ref{subsec:SAO_section}.

\subsection{Missing observations}

In this preferential sampling (i.e. chiefly surveying the areas of
heightened concern, \cite{loperfido_network_2008}) with high screening
costs, not all pollutant concentrations are tracked at each monitoring
site. Out of 35 sites, only C416 (black pin in Figure \ref{fig:site_map})
measured all 5 pollutants. Also, TCEQ uses nearly 30 codes to identify
invalid measurements resulting from downtimes, data losses, rejected
measurements, equipment malfunctions, etc. Our dataset contained 16
such codes, which we consider to be missing data. 

We impute short temporal stretches of NAs, defined as up to 4 contiguous
hourly NAs from the same site within each air pollutant, with monotone
Hermite splines \cite{dougherty_nonnegativity-_1989}. The advantage
of this method is that imputed observations stay within the bounds
of starting and ending observed values, which prevents negative imputations
near extreme observations noted with other methods. A similar approximation
could have been achieved with linear approximates, but we feel that
splines can better incorporate the nearby diurnal structure, if only
a few consecutive observations are missing. 

The larger gaps are replaced by the spatial averages within each pollutant,
when we construct a spatially averaged observations (SAO) indicator
in Section \ref{subsec:SAO_section}. 

The summary of missing values and data imputations are given in Table\ref{tab:Missing_Data_Sumary}.
Apparently, most sites are equipped to gauge ozone, while CO, $\text{SO}_{2}$,
and $\text{PM}_{2.5}$ are quantified at only a handful of locations.
The short NA gaps are least troublesome with $\text{PM}_{2.5}$ and
$\text{O}_{3}$ observations. Notably, $\text{NO}_{2}$ and $\text{PM}_{2.5}$
stand out with larger proportion of missing data.  

\begin{table}[h]
\begin{centering}
\begin{tabular}{lccccc}
\hline 
 & $\text{O}_{3}$ & $\text{CO}$ & $\text{NO}_{2}$ & $\text{SO}_{2}$ & $\text{PM}_{2.5}$\tabularnewline
\hline 
Long runs & 2.89 & 1.90 & 3.16 & 1.94 & 6.30\tabularnewline
Short runs & 0.95 & 1.44 & 2.13 & 1.36 & 0.58\tabularnewline
Total & 3.84 & 3.34 & 3.34 & 3.30 & 6.88\tabularnewline
\#sites & 34 & 7 & 13 & 6 & 5\tabularnewline
\hline 
\end{tabular}
\par\end{centering}
\caption{\label{tab:Missing_Data_Sumary} Proportion of missing values (in
\%) attributed to Long and Short runs Off-line sites (non-contributing
for over one month) are dropped from NA summary for the non-participation
period: C695, C696 for CO and C555, C572, C695, C696 for $\text{O}_{3}$.}
\end{table}

Adjustment for daylight savings time yield little improvement and
we leave details to an appendix of the paper.

\subsection{\label{subsec:SAO_section}Spatially-averaged observations (SAO) }

It is common to spatially average observations from multiple monitoring
sites. While the true average estimator is unknown, a mean-based indicator,
$\bar{x}_{hdp}$, is a popular choice in literature. Here we index
our observations by hour $h=0..23$, by day $d=0..1095$, and by measured
air pollutant $p=1..5$, representing $\text{O}_{3}$, $\text{CO}$,
$\text{NO}_{2}$, $\text{SO}_{2}$, and $\text{PM}_{2.5}$ respectively.
Such equi-weighted measure of centrality assumes homogeneity among
monitoring sites. In this paper we prefer a more robust, median-based,
measure of spatially averaged observations (SAO), $\text{SAO}_{hdp}:=\tilde{x}_{hdp}$,
and the related matrix $\text{SAO}_{h}:=\text{SAO}_{h..}=\left[\tilde{x}_{hdp}\right]_{\forall dp}=\in\mathbb{R}_{+}^{1095\times5}$.

In our PCA median-based $\text{SAO}_{h}$ performs better than its
mean-based counterpart, yielding a clearer cyclostationary EV pattern,
more stable $\text{PC}_{h.1}$ coefficients (see Section \ref{sec:Results-and-discussion}).
Other indicators considered in practice and literature include the
use of a maximum (i.e. aids in study of air pollution peaks and health),
a combination of averaging functions, and a multi-level aggregation,
such as spatial clustering of sites based on some notion of similarity.
Bruno \cite{bruno_unified_2002}, Lee \cite{lee_constructing_2011}
and references therein present a good overview of various air quality
indicators. 

\begin{figure}[h]
\begin{centering}
\includegraphics[width=1\textwidth]{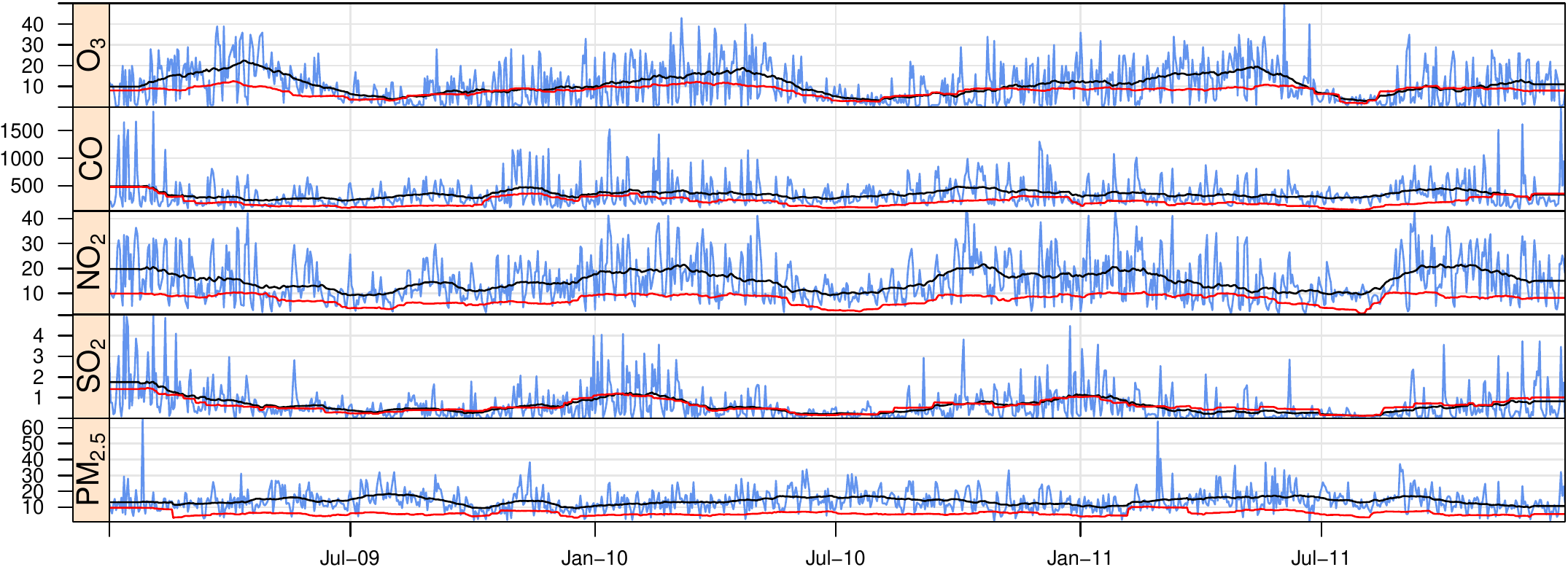}
\par\end{centering}
\caption{\label{fig:SAO} Spatially averaged observations at 7am, $\text{SAO}_{\text{7am}}$.
Measurements ($\mu\text{g}/\text{m}^{3}$ for $\text{PM}_{2.5}$,
ppb for others) are adjusted for DST. Overlaid curves are 45-day rolling
statistics: simple mean (black), standard deviation (red). }
\end{figure}

Raw (non-standardized) SAO are shown in Figure \ref{fig:SAO} along
with rolling ($\ell=45$-day) mean and standard deviation. Note the
non-stationarity of the data expressed with time-dependent mean and
variance. For example, the first two sample moments $\text{NO}_{2}$
and $\text{SO}_{2}$ are elevated in winters, those of $\text{PM}_{2.5}$
- in summers. The clustered behavior persists across all pollutants.
Yet, covariance is more difficult to observe due to dissimilar scale
and embedded noise. 

\subsection{\label{subsec:Normalization-and-standardization}Normalization and
standardization}

\begin{figure}[h]

\begin{minipage}[t]{0.48\columnwidth}%
\subfloat[\label{fig:p-val} Horizon plots showing 45-day rolling p-values,
$p_{7.}$ , of (Henze-Zirkler's) MVN tests on $\text{SAO}_{\text{7}}$,
$\text{LSAO}_{\text{7}}$, and $\text{NSAO}_{\text{7}}$ datasets.
More blue indicates higher likeliness of the underlying data following
MVN, implying fewer outliers, and, thus, greater suitability of PCA.
See \cite{korkmaz_mvn:_2014}.]{\begin{centering}
\includegraphics[width=1\columnwidth]{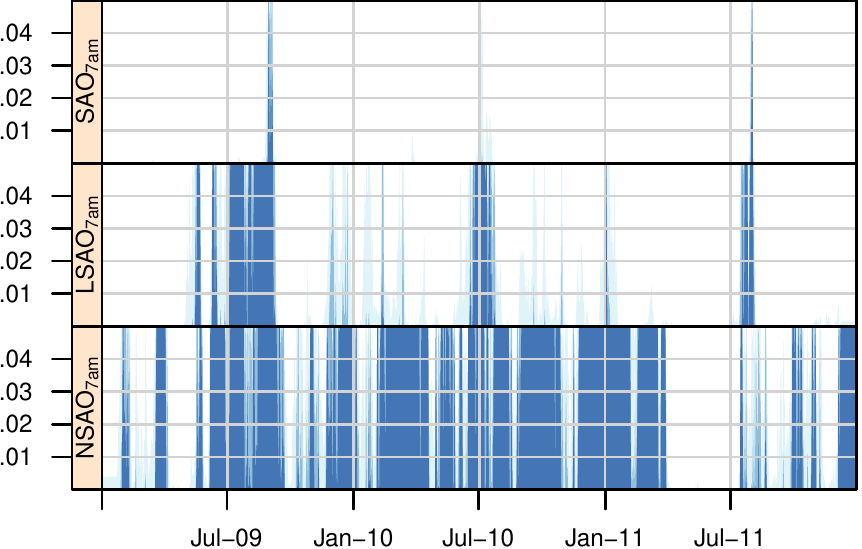}
\par\end{centering}
}%
\end{minipage}\hfill{}%
\begin{minipage}[t]{0.48\columnwidth}%
\subfloat[{\label{fig:outliers} Horizon plots showing 45-day rolling proportions
of outliers for $\text{SAO}_{\text{7}}$, $\text{LSAO}_{\text{7}}$,
and $\text{NSAO}_{\text{7}}$ datasets. Detection is based on adjusted
robust Mahalanobis distance, $\text{rMD}\left(\cdot\right)$, with
decision on parameter $\alpha\in\left[.5,1\right]$ (we use $\alpha=.75$).
We note a consistent relative outcome: $\text{SAO}_{\text{7}}$ contains
most outliers (more blue) $\text{NSAO}_{\text{7}}$ has least. See
\cite{korkmaz_mvn:_2014}}]{\begin{centering}
\includegraphics[width=1\columnwidth]{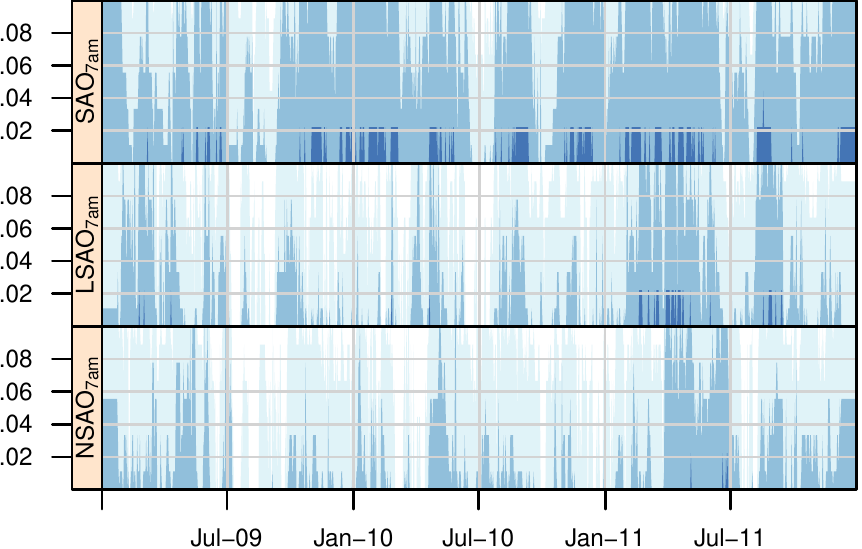}
\par\end{centering}
}%
\end{minipage}
\end{figure}

As part of robustifying $\text{SAO}_{h}$, we have assessed logarithmic
and other non-linear transformations, which are common in the examination
of AQ data. Since normalization (herein aligning data to MVN) is usually
associated with robustifying PCA (see Section \ref{subsec:Robustness}),
it is reasonable to use an MVN test to target the desirable transform.
The 45-day moving window p-values, $p_{hd}$, of Henze-Zirkler's MVN
test are presented in Figure \ref{fig:p-val}. That is $p_{h.}\in\mathbb{R}^{1095}$
is a non-local time series of (daily) p-values fixed at $h\in\left\{ 0..23\right\} $,
hour of a day. The plot has a 5\% significance level cut off; and,
more blue indicates a greater likelihood of tested data following
MVN.  The summary observations from Figure \ref{fig:p-val} are:
\begin{enumerate}
\item The top panel shows that non-transformed data, $\text{SAO}_{\text{7am}}$,
fails to exhibit normality at 7am. This time of a day is representative
of daily traffic build up.
\item The middle panel reflects a slight improvement in MVN test of log
transform 
\begin{eqnarray*}
\hat{x}_{hdp} & := & \log\left(1+\tilde{x}_{hdp}\right)\\
\text{LSAO}_{h} & := & \hat{x}_{h..}\in\mathbb{R}^{1095\times5}
\end{eqnarray*}
\end{enumerate}
where $\tilde{x}_{hdp}$ is defined in Section \ref{subsec:SAO_section}.
\begin{enumerate}
\item The bottom panel illustrates log differencing as a considerably promising
normalization. It is a routine method in financial models, which use
log returns, or percent change, computed analogously from the observed
stock prices. Similarly, we define normalized SAO as 
\begin{eqnarray}
y_{hdp} & := & \hat{x}_{hdp}-\hat{x}_{h,d-1,p}\label{eq:Diff-Log}\\
\text{NSAO}_{h} & := & y_{h..}\in\mathbb{R}^{1095\times5}
\end{eqnarray}
\end{enumerate}
Also, as expected, median-based $\text{SAO}_{h}$ exhibits greater
normality than a similar mean-based measure across the evaluated transformations.
Other MVN tests (see Section \ref{subsec:Robustness}) also support
the use of the median-based transform defined in \eqref{eq:Diff-Log}.
Likewise, other transforms listed in Section \ref{subsec:Robustness}
yield similar-to-slightly-inferior performance as that of log mapping
($\text{LSAO}_{h}$). When $\text{NSAO}_{h}$ is assessed at other
hours of a day (night time, traffic time, etc.), the MVN test's conclusions
are similar. 

Outliers, assessed for the same three transformations, are presented
in Figure \ref{fig:outliers} and also support the use of use of log
differencing. Hence, we proceed with the analysis on median-based
$\text{NSAO}_{h}$ data.

Raw concentrations use different scales and are not suited for PCA,
as noted in Section \ref{subsec:Definition-and-interpretation}. A
common approach is to standardize the units to have mean 0 and variance
1 prior to application of PCA \cite{buhr_assessment_1992,wold_principal_1987}.
We do so on each 45-day window. For instance, CO measurements dominate
the results of PCA of $\text{SAO}_{h.2}$, if left unscaled (see data
summary in Table \ref{tab:Summary_Stats}). Similarly, non-standardized
$\text{O}_{3}$ observations govern PCA of  LSAO and NSAO because
its variability is up to twice that of other variables. In fact, when
PCA was tried on unscaled $\text{NSAO}_{7}$, $\text{PC}_{7.1}$ explained
80\% of variability with dynamic loadings for the (normalized) $\text{O}_{3}$
quantities playing a prominent part. In contrast, normalized $\text{O}_{3}$
participation is similar to that of $\text{CO}$ and $\text{NO}_{2}$
in $\text{PC}_{7.1}$, when $\text{NSAO}_{7}$ is standardized. Since
PCs are designed to capture and attribute variables' variability,
the former $\text{PC}_{7.1}$ is likely inflated by variability of
$\text{O}_{3}$.

Table \ref{tab:Summary_Stats} describes raw, log and log differenced
pollutant indicators. Note that NSAO variable's mean and median are
nearly identical, an expected property of data from MVN distribution.
While SAO and LSAO exhibit dramatic differences in various statistic
measures (across pollutants), NSAO pollutants' statistics (min, max,
...) are better aligned. In our analysis we do not require strict
normality. Our primary goal is to prepare data for DPCA by minimizing
the effect of outliers on each rolling subsample.    

\begin{table}[h]
\begin{centering}
{\footnotesize{}}%
\begin{tabular}{r|rrrrr|rrrrr|rrrrr}
\multicolumn{1}{r}{} & \multicolumn{5}{c|}{{\scriptsize{}SAO...}} & \multicolumn{5}{c|}{{\scriptsize{}LSAO...}} & \multicolumn{5}{c}{{\scriptsize{}NSAO...}}\tabularnewline
\cline{2-16} 
 & {\scriptsize{}O3} & {\scriptsize{}CO} & {\scriptsize{}NO2} & {\scriptsize{}SO2} & {\scriptsize{}PM2.5} & {\scriptsize{}O3} & {\scriptsize{}CO} & {\scriptsize{}NO2} & {\scriptsize{}SO2} & {\scriptsize{}PM2.5} & {\scriptsize{}O3} & {\scriptsize{}CO} & {\scriptsize{}NO2} & {\scriptsize{}SO2} & {\scriptsize{}PM2.5}\tabularnewline
\cline{2-16} 
{\scriptsize{}Min} & {\scriptsize{}0} & {\scriptsize{}6.64} & {\scriptsize{}.82} & {\scriptsize{}0} & {\scriptsize{}.10} & {\scriptsize{}0} & {\scriptsize{}2.03} & {\scriptsize{}.60} & {\scriptsize{}0} & {\scriptsize{}.09} & {\scriptsize{}-1.73} & {\scriptsize{}-1.62} & {\scriptsize{}-1.08} & {\scriptsize{}-1.30} & {\scriptsize{}-2.20}\tabularnewline
{\scriptsize{}1Q} & {\scriptsize{}13.00} & {\scriptsize{}142.33} & {\scriptsize{}4.64} & {\scriptsize{}.12} & {\scriptsize{}7.27} & {\scriptsize{}2.64} & {\scriptsize{}4.97} & {\scriptsize{}1.73} & {\scriptsize{}.12} & {\scriptsize{}2.11} & {\scriptsize{}-0.11} & {\scriptsize{}-.09} & {\scriptsize{}-.11} & {\scriptsize{}-.07} & {\scriptsize{}-.10}\tabularnewline
{\scriptsize{}Med} & {\scriptsize{}23.00} & {\scriptsize{}182.61} & {\scriptsize{}7.26} & {\scriptsize{}.33} & {\scriptsize{}10.23} & {\scriptsize{}3.18} & {\scriptsize{}5.21} & {\scriptsize{}2.11} & {\scriptsize{}.29} & {\scriptsize{}2.42} & {\scriptsize{}0} & {\scriptsize{}0} & {\scriptsize{}0} & {\scriptsize{}0} & {\scriptsize{}.01}\tabularnewline
{\scriptsize{}Mean} & {\scriptsize{}24.83} & {\scriptsize{}220.67} & {\scriptsize{}9.71} & {\scriptsize{}.60} & {\scriptsize{}11.30} & {\scriptsize{}2.99} & {\scriptsize{}5.26} & {\scriptsize{}2.18} & {\scriptsize{}.39} & {\scriptsize{}2.40} & {\scriptsize{}0} & {\scriptsize{}0} & {\scriptsize{}0} & {\scriptsize{}0} & {\scriptsize{}0}\tabularnewline
{\scriptsize{}3Q} & {\scriptsize{}34.00} & {\scriptsize{}242.26} & {\scriptsize{}12.13} & {\scriptsize{}.74} & {\scriptsize{}14.21} & {\scriptsize{}3.56} & {\scriptsize{}5.49} & {\scriptsize{}2.57} & {\scriptsize{}.55} & {\scriptsize{}2.72} & {\scriptsize{}.07} & {\scriptsize{}.09} & {\scriptsize{}.11} & {\scriptsize{}.05} & {\scriptsize{}.11}\tabularnewline
{\scriptsize{}Max} & {\scriptsize{}101.00} & {\scriptsize{}2076.51} & {\scriptsize{}50.52} & {\scriptsize{}19.75} & {\scriptsize{}81.35} & {\scriptsize{}4.62} & {\scriptsize{}7.64} & {\scriptsize{}3.94} & {\scriptsize{}3.03} & {\scriptsize{}4.41} & {\scriptsize{}2.40} & {\scriptsize{}1.54} & {\scriptsize{}1.17} & {\scriptsize{}1.44} & {\scriptsize{}1.59}\tabularnewline
{\scriptsize{}SD} & {\scriptsize{}15.68} & {\scriptsize{}149.62} & {\scriptsize{}7.35} & {\scriptsize{}.85} & {\scriptsize{}5.84} & {\scriptsize{}.85} & {\scriptsize{}.49} & {\scriptsize{}.60} & {\scriptsize{}.37} & {\scriptsize{}.48} & {\scriptsize{}.29} & {\scriptsize{}.19} & {\scriptsize{}.21} & {\scriptsize{}.16} & {\scriptsize{}.22}\tabularnewline
\end{tabular}
\par\end{centering}{\footnotesize \par}
\caption{\label{tab:Summary_Stats} Summary statistics for $\text{SAO}$, $\text{LSAO}$
and $\text{NSAO}$}
\end{table}

\subsection{Dynamic correlation }

\begin{figure}[h]
\includegraphics[width=0.49\textwidth]{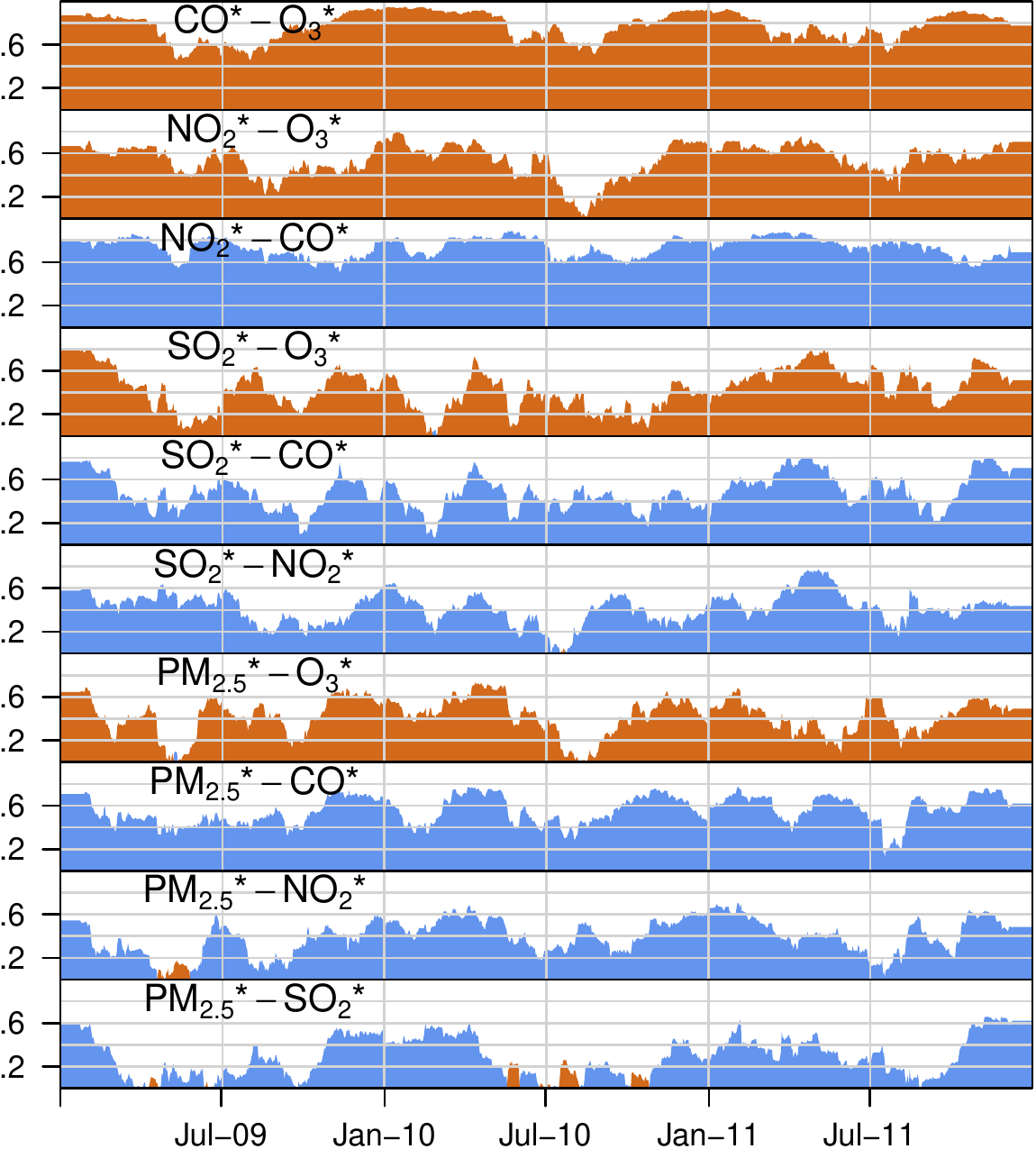}\hfill{}\includegraphics[width=0.49\textwidth]{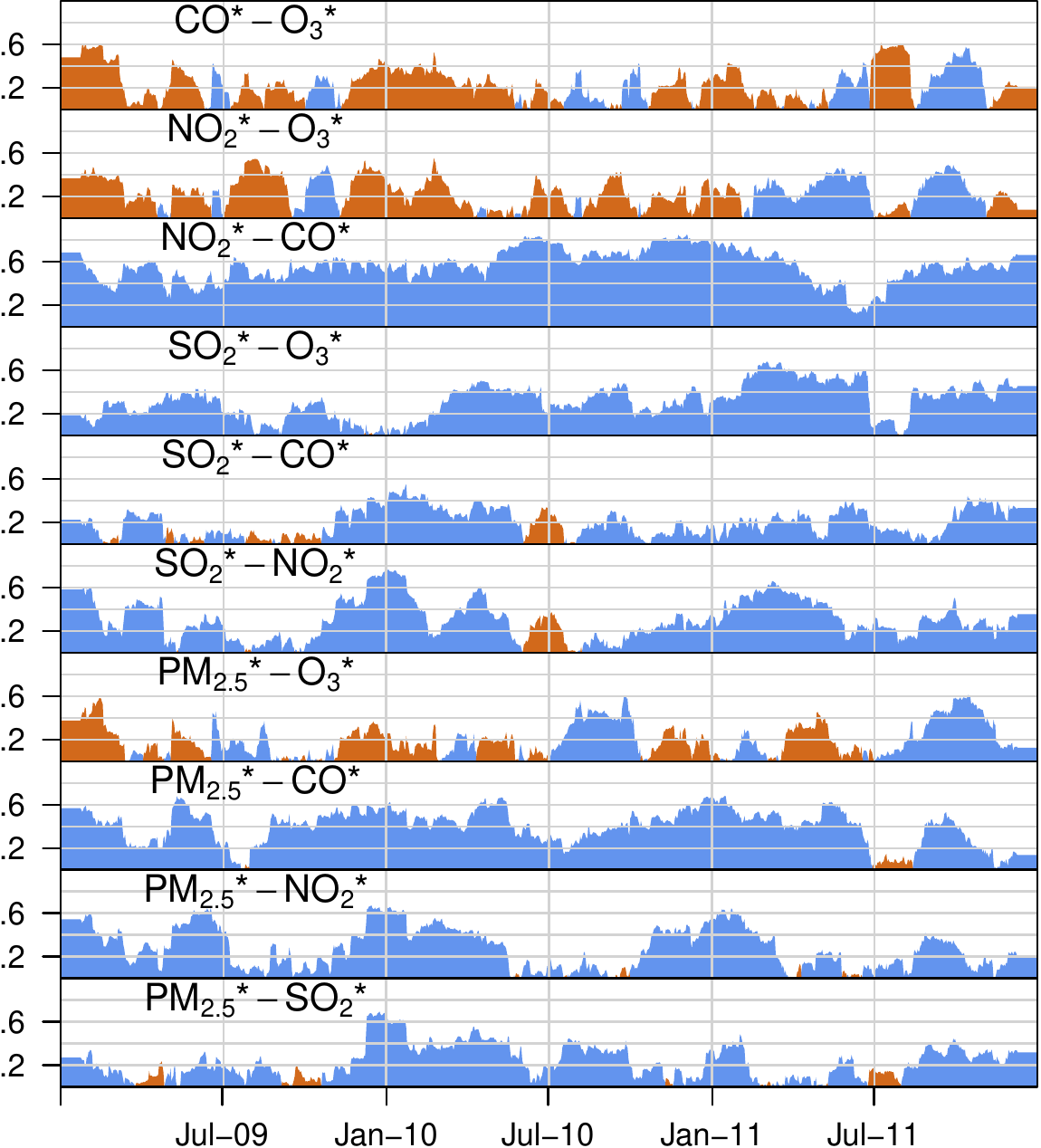}

\caption{\label{fig:Corr} Pearson correlations for the paired $\text{NSAO}_{h}$
variables are computed on a 45-day rolling window in the morning (7am,
\emph{left}) and afternoon (2pm, \emph{right}). Positive/negative
correlations are shown in blue/red, both on a positive axis. }
\end{figure}

PCA maps highly correlated variables to uncorrelated components. It
would make little sense to apply PCA to uncorrelated variables. So,
we quickly check the degree of association between normalized pollutants.
Indeed, as shown in Figure \ref{fig:Corr}, some variables of $\text{NSAO}_{h\in\left\{ 7,14\right\} }$
exhibit a high degree of contemporaneous dependency. $\text{O}_{3}^{*}$
are strongly associated with $\text{CO}^{*}$ and $\text{NO}_{2}^{*}$
in the morning, but not in the afternoon (where {*} indicates a normalized
observation). $\text{NO}_{2}^{*}$ is correlated with $\text{CO}^{*}$
in both samples. 

In general, morning correlations are more substantial than those in
the afternoon. Also, a seasonal pattern is observable in some correlations.
For example, morning $\text{CO}^{*}$ to $\text{O}_{3}^{*}$ correlations
are more negative in the winters and than in the summers. Thus, we
have established that the issue of co-dependence is significant and
the use of PCA is just. Also, the presence of seasonal cycles underlines
the cyclostationary structure of the data and supports the use of
DPCA.

\subsection{Choice of a window size}

Air pollution data carries clear seasonal and diurnal patterns. Its
cyclostationarity allows us to assume a fixed mean and variance on
a short (length $\ell$) window of observations. We assume that $\ell=45$
days carries sufficient information to grasp the approximately stationary
structure at a particular time of a year. 

\section{\label{sec:Results-and-discussion}Results and discussion}

\subsection{\label{subsec:EV}Explained variance (EV)}

\begin{figure}[h]
\includegraphics[width=0.327\textwidth]{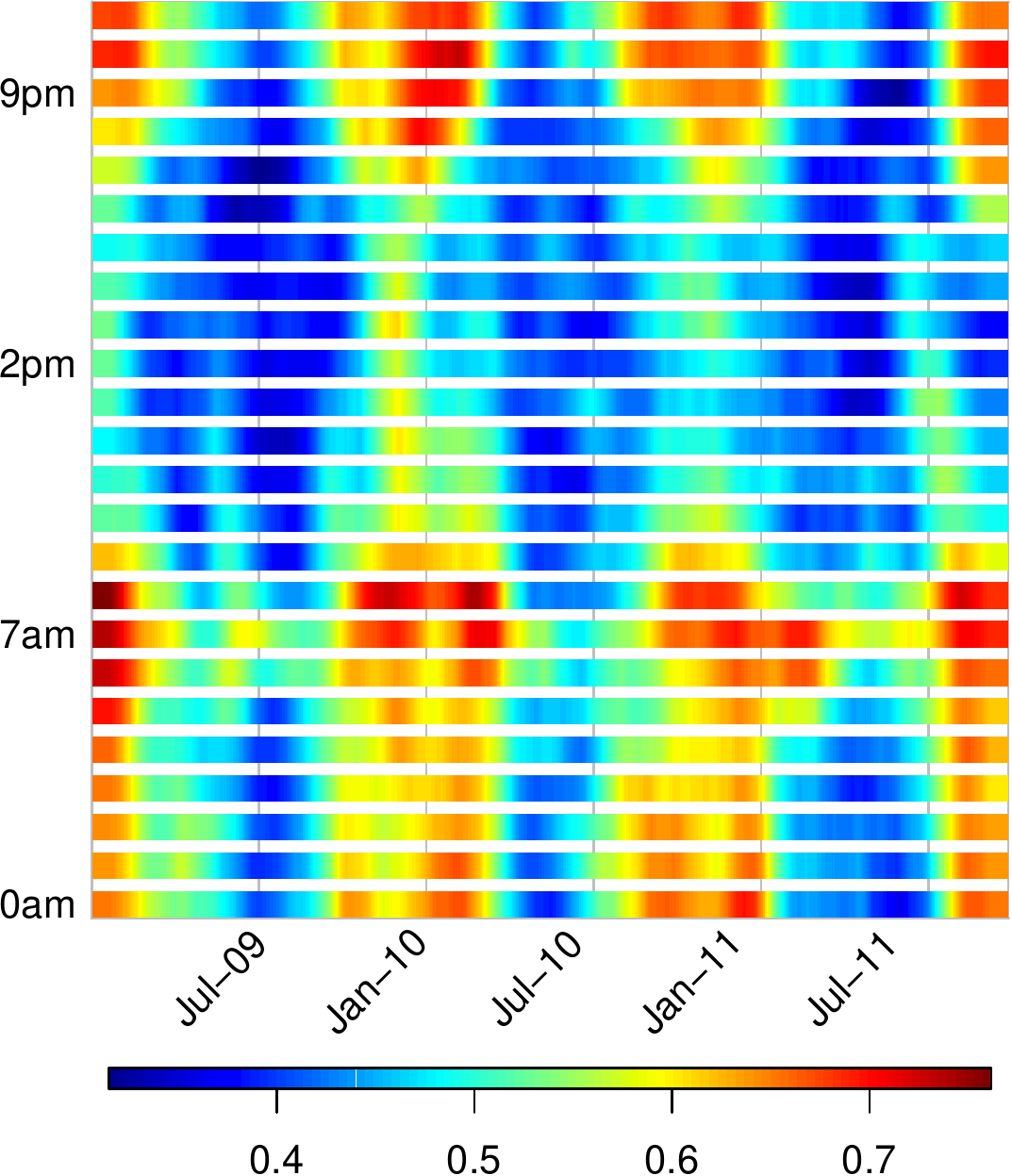}\hfill{}\includegraphics[width=0.327\textwidth]{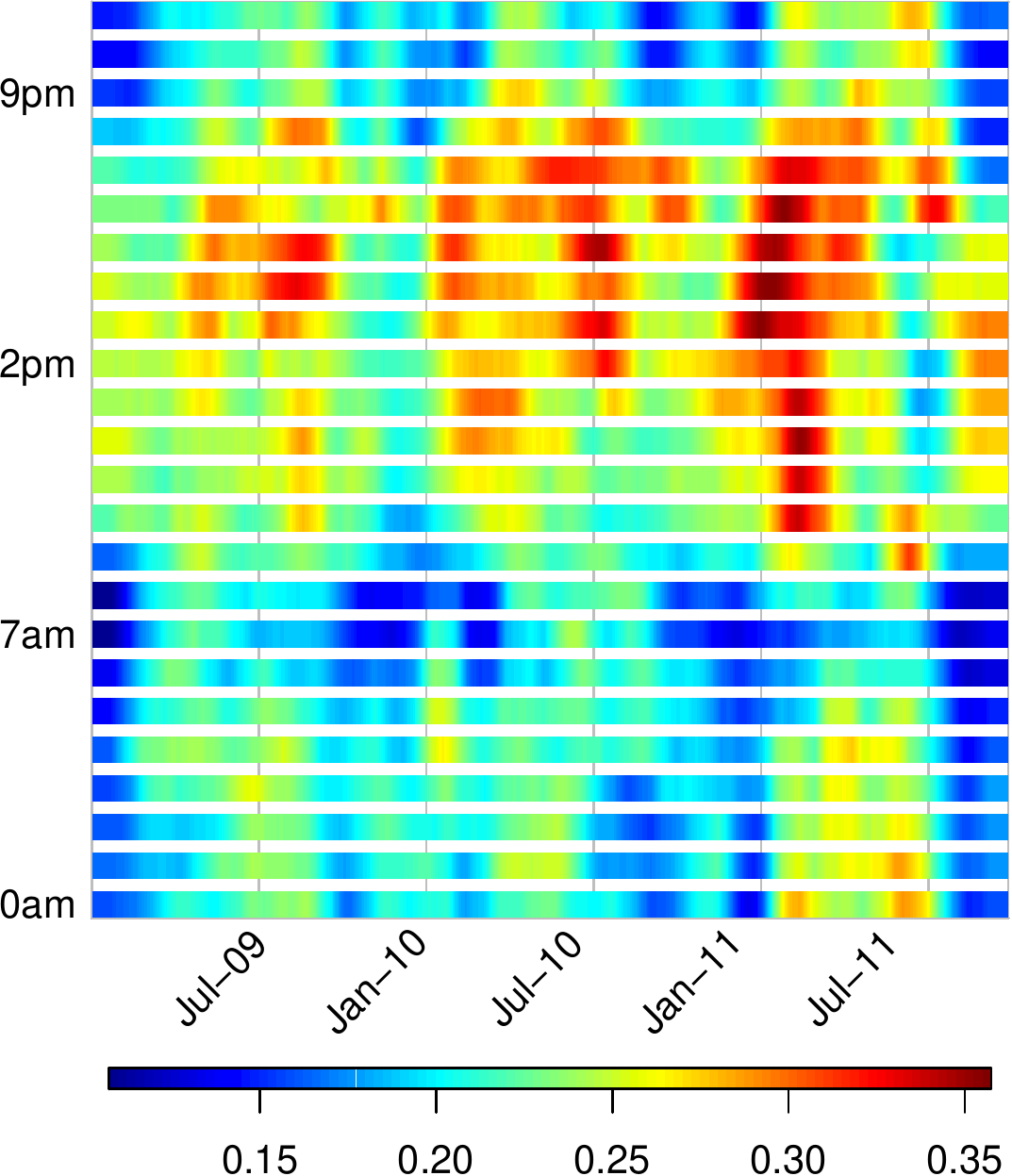}\hfill{}\includegraphics[width=0.327\textwidth]{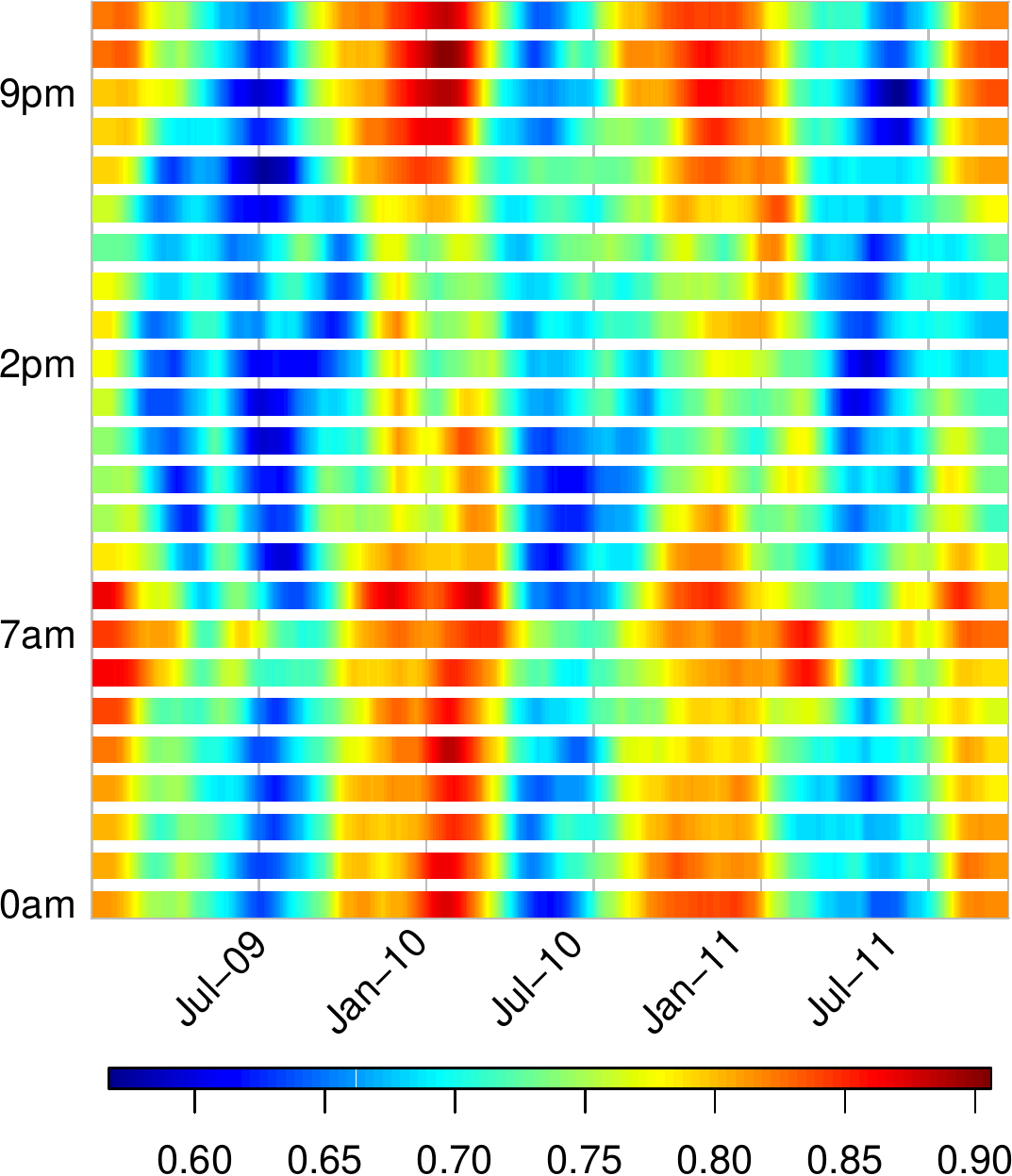}

\caption{\label{fig:CEV-2D-NSAO} Heatmaps of the dynamic explained variance
(EV) from $\text{PC}_{h.1}$, $\text{PC}_{h.2}$, and $\text{PC}_{h.1}+\text{PC}_{h.2}$
of the $\text{NSAO}_{h},h=0..23$, \emph{left}, \emph{right} and \emph{center},
respectively.}
\end{figure}

Examination of the dynamic nature of the explained variance for $\text{PC}_{h.1}$,
$\text{PC}_{h.2}$ and the pointwise sum of the two for the $\text{NSAO}_{h}$,
at each hour of a day, over the three-year study period yields key
insights. Figure \ref{fig:CEV-2D-NSAO} depicts these components.
In general, higher explained variance corresponds to a better PCA
fit and stronger linear relations among PCA input variables that make
up the PCs. 

We use \textbf{R} (version 3.x) core (\texttt{base}, \texttt{stats}),
\texttt{xts} and \texttt{lattice} packages for most of data scrubbing,
imputation, PCA and visualization. Non-local $\text{NSAO}_{h}$ are
standardized on each 45-day window before PCA is applied and Figure
\ref{fig:CEV-2D-NSAO} of dynamic EV is drawn. This 3D plot profiles
EV components over a 2D time domain as a non-local (daily) pattern
of $\text{EV}_{h.k}$ and local (hourly) pattern of $\text{EV}_{.dk}$
. 

Admirably, just two PCs explain up to 90\% of variability in the components
(in morning winters). But, more importantly, such profiling presents
the EV pattern of the components ($\text{PC}_{..k\in\left\{ 1,2\right\} }$)
dissected by time of day and day of the observed period. 

The daily explained variability by the first principal component at
hour $h$, $\text{EV}_{h.1}$, exhibits a strong seasonal trend, spiking
in cool winters and sinking in hot and humid Texas summers, for any
fixed hour of a day. A trend non-stationary $\text{EV}_{..1}$ ranges
from about 30\% to about 75\% with overall mean, $\overline{\overline{\text{EV}}}_{1}$,
of approximately $51\%$ as shown in Table \ref{tab:Total_mean_EV}. 

The seasonal form of $\text{CEV}_{h.2}$ follows that of $\text{EV}_{h.1}$
because the marginal difference, i.e. $\text{EV}_{h.2}$, is relatively
too small and less variable. The mean of $\text{EV}_{h.2}$ is less
than half of the mean of $\text{EV}_{h.1}$ (23\% vs 51\%, see Table
\ref{tab:Total_mean_EV}). Overall mean variance explained by the
first two principal components is $\overline{\overline{\text{CEV}}}_{2}\approx74\%$. 

The measure $\text{EV}_{.d1}$ exhibits a strong diurnal pattern,
when the figure panels are assessed vertically with changing hours
of a day. The contributions are higher overnight, from late evening
to early morning, peaking with sun rise at around 7am. These times
of a day exhibit very little direct solar radiation. Contributions
drop in the afternoons, reaching lowest points around 4-5 pm. Such
diurnal pattern is strongest in the winters. Diurnal contributions
from the second component, $\text{EV}_{.d2}$, slightly smooth out
this diurnal pattern with elevated contribution mid-day and lower
contributions at night. As a result, the patterns are less prominent
in the right panel showing $\text{CEV}_{h.2}$. 

Naturally, static $\text{EV}_{k}$ fails to capture such complex diurnal
and cyclostationary dynamics.

\subsection{\label{subsec:Non-local-mean-of-EV}Mean explained variance}

Eyeballing 3D EV (\prettyref{fig:CEV-2D-NSAO}) is helpful as it reveals
a great deal of detail. However, for a quick assessment of intraday
contribution behavior, one may consider non-locally averaged EV, computed
at a specific hour as 

\begin{eqnarray}
\overline{\text{EV}}_{hk} & := & \frac{1}{\mathfrak{d}}\sum_{d}\text{EV}_{hdk}\label{eq:Mean_EV}\\
\overline{\text{CEV}}_{hk} & := & \frac{1}{\mathfrak{d}}\sum_{i\le k}\overline{\text{EV}}_{hi}\nonumber 
\end{eqnarray}
where number of days $\mathfrak{d}=1095$.

The plots in these section focus on analysis of quantities in \eqref{eq:Mean_EV}
and their (somewhat limited due to aggregation) use as a measure of
PCA performance.

\begin{figure}[H]
\begin{centering}
\includegraphics[width=0.5\columnwidth]{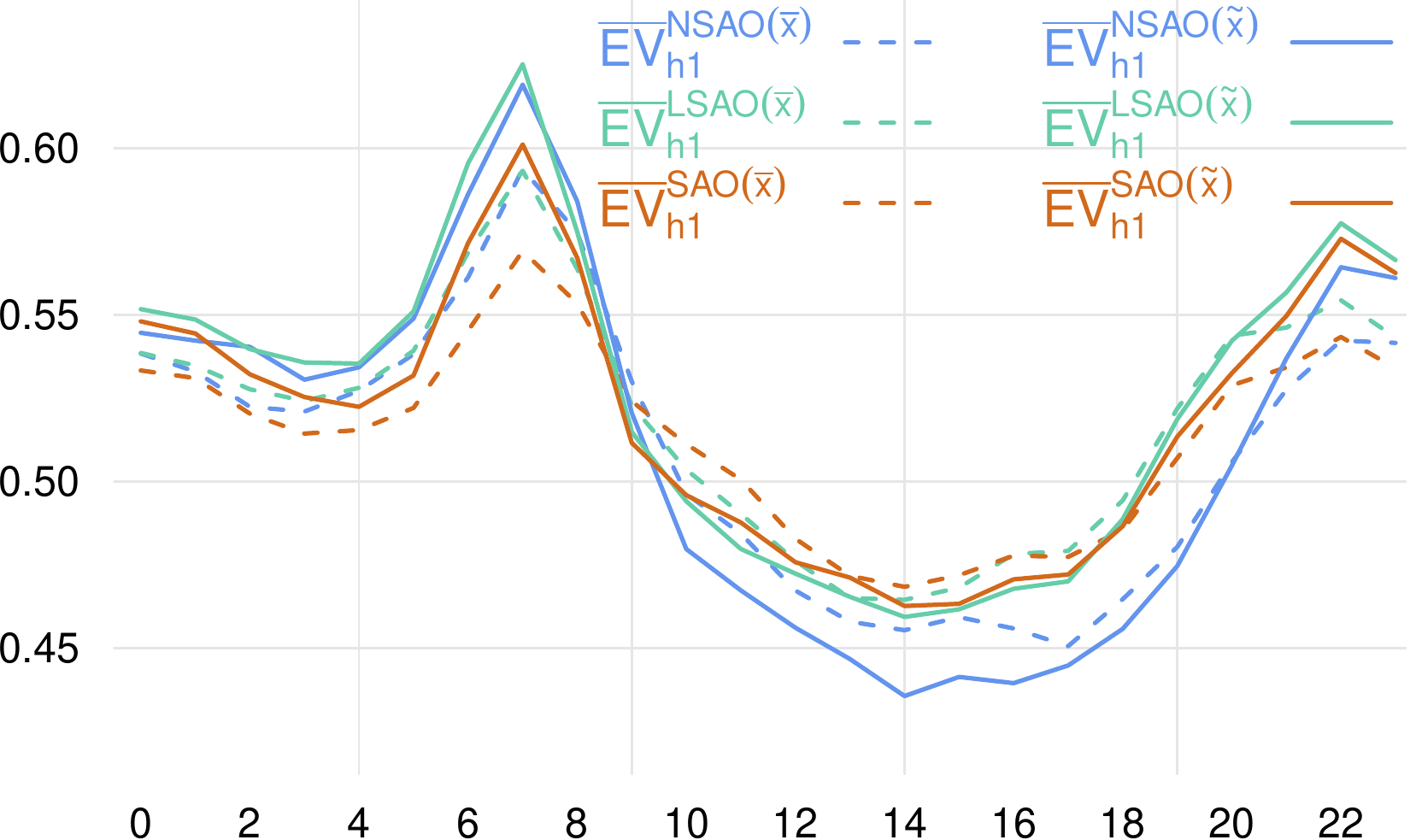}
\par\end{centering}
\caption{\label{fig:EV_means_compare} Comparison of $\overline{\text{EV}}_{hk}$.
Dotted lines use $\text{SAO}_{h}$ based on spatial mean ($\bar{x}_{hdp}$),
i.e. averaging of contemporaneous observations among monitoring sites.
Solid lines uses $\text{SAO}_{h}$ based on spatial median ($\tilde{x}_{hdp}$).
Blue line computes $\overline{\text{EV}}_{hk}$ based on $\text{NSAO}_{h}$,
green - $\text{LSAO}_{h}$, brown - $\text{SAO}_{h}$. Ordinate units
are in proportions, abscissa - in hours of a day (0 to 23).}
\end{figure}

To start off, we want to evaluate our choice of $\text{SAO}$ averaging
function and normalizing transformation. We briefly consider Figure
\ref{fig:EV_means_compare} for such comparison. It exhibits $\overline{\text{EV}}_{hk}$
based on $\text{SAO}_{h}$ (identity transform), $\text{LSAO}_{h}$
(log transform), and $\text{NSAO}_{h}$ (log differencing transform),
where input $\text{SAO}_{h}$ is computed either via mean or median
function, i.e. $\bar{x}_{hdp}$ and $\tilde{x}_{hdp}$, respectively
(see Section \ref{subsec:SAO_section}). The overall shapes appear
similar across all spatial averaging and normalizing methods. That
is $\overline{\text{EV}}_{hk}$ spikes at 7am and dips in the afternoon
(1-5pm). Thus, at least with the $\overline{\text{EV}}_{h1}$ measure,
these methods do not grossly differ at representing the aggregate
dynamics of underlying variables. Still $\bar{x}_{hdp}$ performs
poorer (vs. $\tilde{x}_{hdp}$) around a peak (7am) and performs vaguely
better in the afternoon (the bottom of the curve). Also, $\text{LSAO}_{h}$
and $\text{NSAO}_{h}$ of $\tilde{x}_{hdp}$ perform best near peak,
but the former beats the latter at most hours of a day. If this aggregate
was a single measure of performance of PCA analysis, then we would
perform PCA on $\text{LSAO}_{h}$, as it is frequently done. However,
the consideration of robustness in Figure \ref{fig:outliers} demands
for PCA on $\text{NSAO}_{h}$, which produces clearer DPCA components.
That is dynamic $\text{EV}_{h.1}$ possess a coherent seasonal structure
in Figure \ref{fig:CEV-2D-NSAO} and dynamic loadings in Figure \ref{subsec:Dynamic-loadings}
are more interpretable, as compared to those of $\text{LSAO}_{h}$,
whose EV plots we added to supplemented material. 

\begin{figure}[h]
\includegraphics[width=0.49\textwidth]{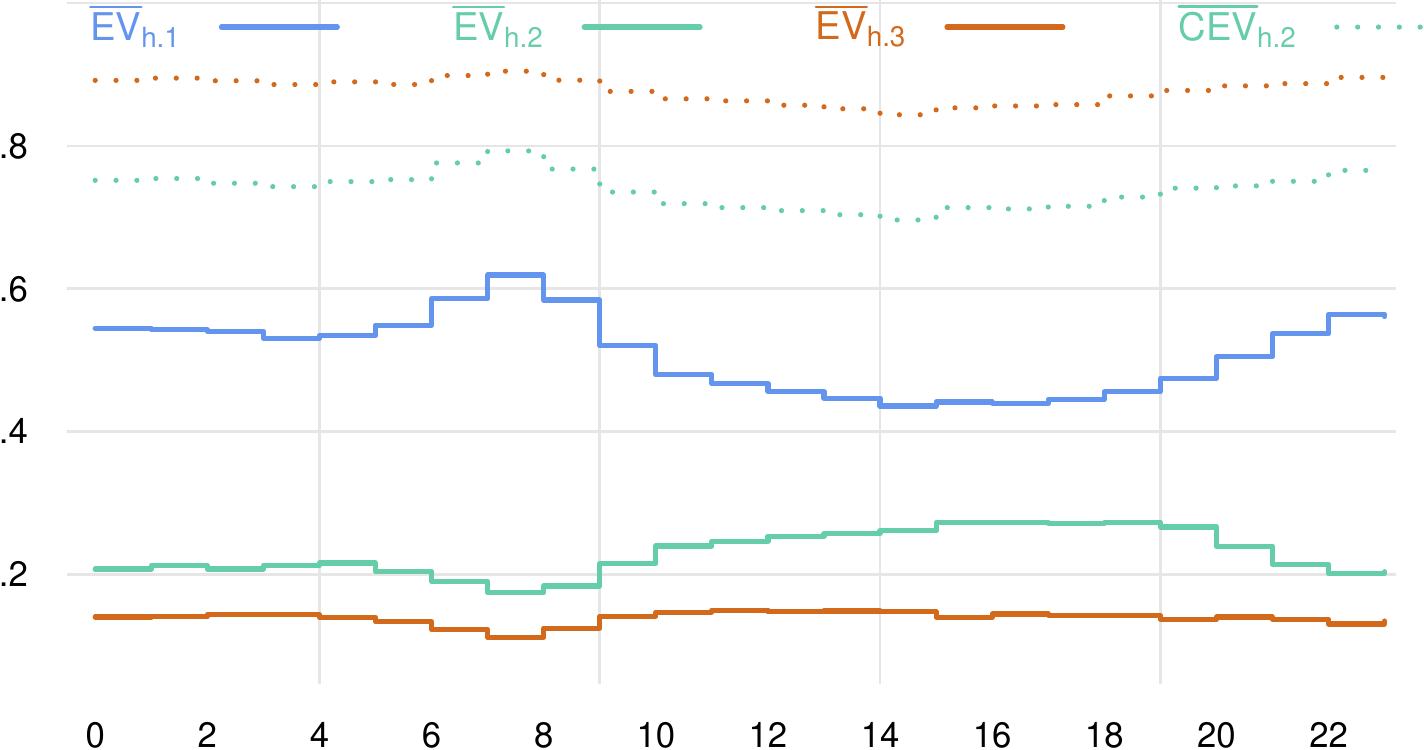}\hfill{}\includegraphics[width=0.49\textwidth]{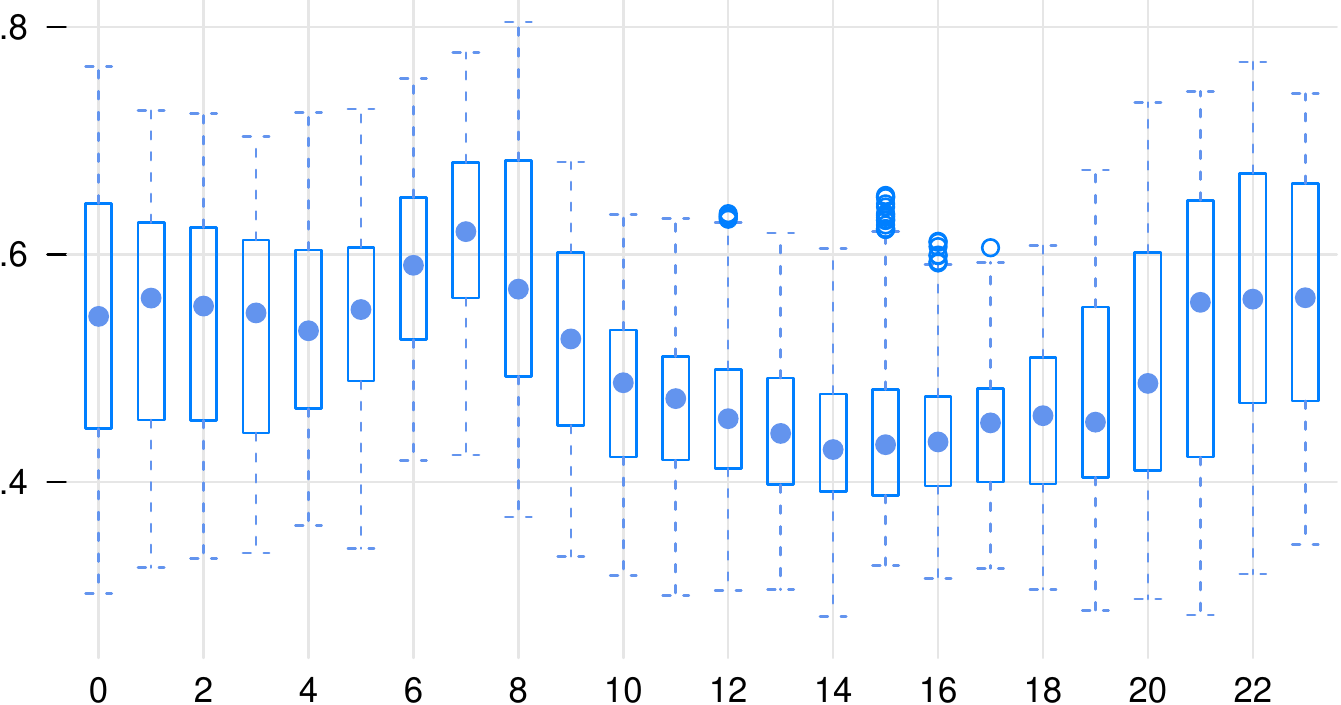}

\caption{\label{fig:EV_means} \emph{Left}: Individual (solid) and cumulative
(dotted) EV for the $\text{PC}_{1}$, $\text{PC}_{2}$, $\text{PC}_{3}$
by hour. \emph{Right}: Distribution of $\text{EV}_{h.1}$ over 3 year
period (ignoring non-stationarity). The centered bullet dots are 24
medians of the underlying samples. Note clustering of outliers near
hours of poor explanatory power (low $\overline{\text{EV}}_{h1}$
values). Ordinate units are in proportions, abscissa - in hours of
a day (0 to 23).}
\end{figure}

We now return to examination of $\overline{\text{EV}}_{hk}$ from
PCA of $\text{NSAO}_{h}$ based on $\tilde{x}_{hdp}$. Figure \ref{fig:EV_means}
reveals the relation between first three $\overline{\text{EV}}_{hk}$
variables ($k=1..3$, $h=0..23$). It shows that $\overline{\text{EV}}_{h1}$
is negatively correlated with $\overline{\text{EV}}_{h2}$. So, when
$\text{PC}_{h.1}$ gains prominence in capturing variability (around
6-8am and midnight), $\text{PC}_{h.2}$ gives up almost as much, and
vise versa. The average explanatory power exceeds 60\% at 7am and
dives just below 45\% in the afternoon (2-6pm). 

The box-and-whisker plot is a compact way to describe a sample variability
or its distribution's shape. These (static) descriptions are illustrated
in Figure \ref{fig:EV_means} for $\text{EV}_{h.1}$ at each hour
$h$. Greater number of outliers appear to coincide with poorer performance
of DPCA (in terms of explained variability) around afternoon hours.
Recall (from Figure \ref{fig:CEV-2D-NSAO}) that afternoon hours were
also blurring the seasonality in $\text{EV}_{h.1}$ . 

Note that $\overline{\text{EV}}_{hk}$ oversimplifies the results.
It favors a clearer (``big picture'') diurnal dynamics, while hides
the seasonal structure of the underlying $\text{EV}_{h.k}$. Still,
the plots support the superiority of DPCA in the morning and near-midnight
$\text{NSAO}_{h}$ and inferiority of such analysis on data in the
afternoon hours. If cyclostationarity of $\text{EV}_{h.k}$ needs
to be explicitly exemplified, then boxplots can be assessed on a windowed
time interval (of, say, 45 days). 

Further aggregation along the dimension of day hours is exhibited
in Table \ref{tab:Total_mean_EV}. We compare these $\overline{\overline{\text{EV}}}_{k}$
values to what other authors have achieved with static PCA, in Section
\ref{subsec:Comparison-to-previous-work}. 

\begin{table}[h]
\begin{centering}
\begin{tabular}{c|c|c|c|c}
$\overline{\overline{\text{EV}}}_{1}$ & $\overline{\overline{\text{EV}}}_{2}$ & $\overline{\overline{\text{EV}}}_{3}$ & $\overline{\overline{\text{CEV}}}_{2}$ & $\overline{\overline{\text{CEV}}}_{3}$\tabularnewline
\hline 
.51 & .23 & .14 & .74 & .88\tabularnewline
\end{tabular}
\par\end{centering}
\caption{\label{tab:Total_mean_EV}Cumulative and non-cumulative overall mean
explained variance, i.e. $\overline{\overline{\text{EV}}}_{k}=\frac{1}{24}\sum_{h}\overline{\text{EV}}_{hk}$.}
\end{table}

\subsection{\label{subsec:Dynamic-loadings}Dynamic loading coefficients}

\begin{figure}[h]
\includegraphics[width=0.49\textwidth]{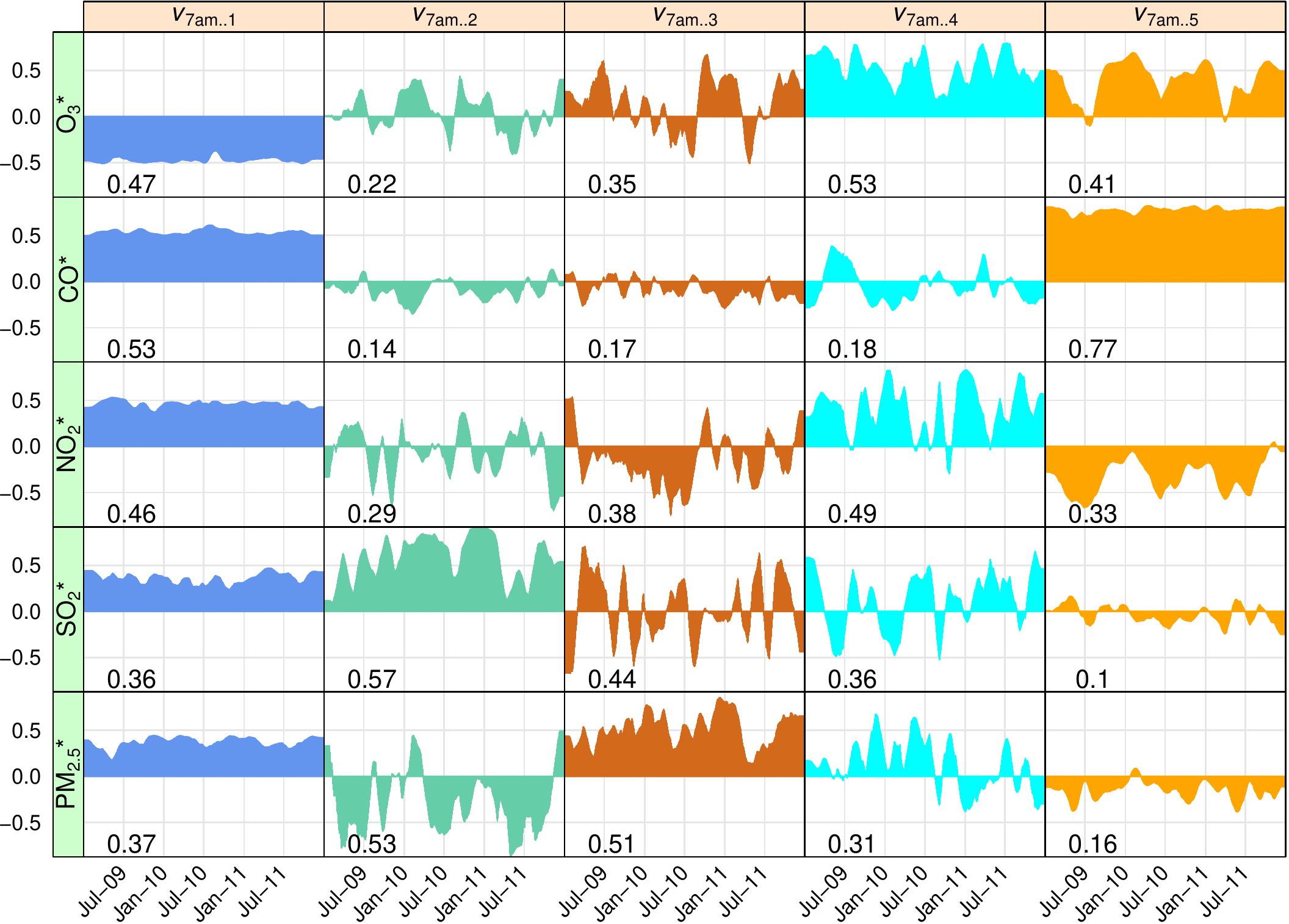}\hfill{}\includegraphics[width=0.49\textwidth]{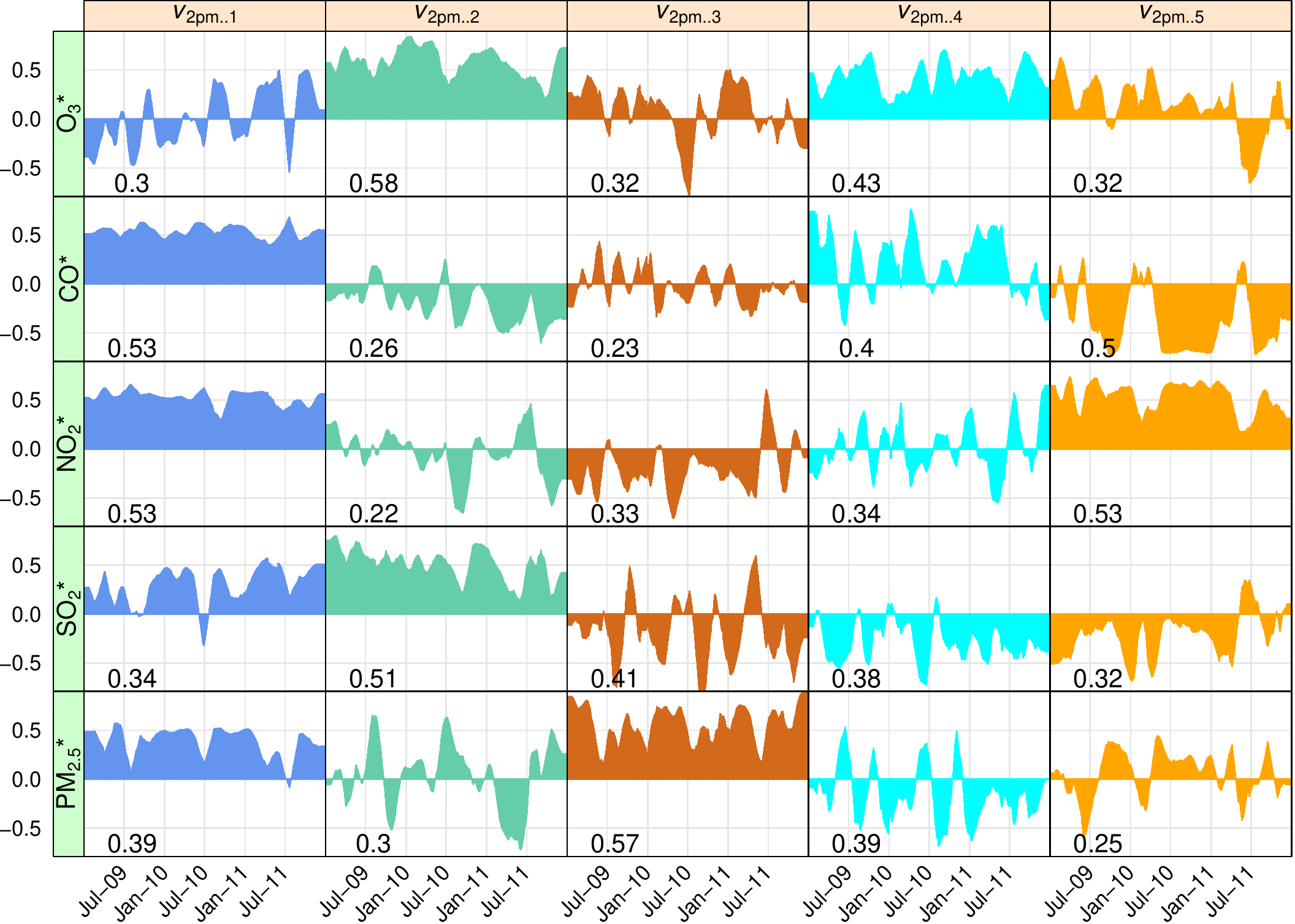}

\caption{\label{fig:PC Loadings}Dynamic loadings at 7am (\emph{left}) and
2pm (\emph{right}). \emph{}}
\end{figure}

Furthermore, we scrutinize the linearity of relationships and participation
of $\text{NSAO}_{h}$ variables (i.e. percent change in pollutants)
in PCs. The two most remarkable hours of a day are 7am and 2pm (see
Figure \ref{fig:EV_means}), when $\overline{\text{EV}}_{hk}$ reaches
its highest and lowest values, respectively. Figure \ref{fig:PC Loadings}
depicts corresponding PC loadings for $h=7,14$. 

From the figure we observe that in the morning  $\text{PC}_{\text{7am}.1}$
(i.e. first dynamic PC for $\text{NSAO}_{\text{7am}}$) is a fairly
consistent linear function of all 5 variables with weights maintaining
their approximate mean and relation to other variables. $\text{CO}^{*}$
(i.e. normalized and standardized $\text{CO}$) is the largest driver
behind $\text{PC}_{7.1}$ with weights averaging 0.53 and reaching
0.6 in summer 2010. Coefficients appear somewhat seasonal with $\text{CO}^{*}$
playing a bigger part of $\text{PC}_{7.1}$ in hot summers. $\text{O}_{3}^{*}$
and $\text{NO}_{2}^{*}$ are also influential. $\text{O}_{3}^{*}$
weights oppose those of all other variables, implying inverse relationship
between log increments of $\text{O}_{3}$ and other pollutants.

Largest (yet unstable) contribution to $\text{PC}_{7.2}$ comes from
$\text{SO}_{2}^{*}$. $\text{PM}_{2.5}^{*}$, second largest, has
opposite sign weights, implying offsetting contribution to $\text{PC}_{7.2}$.
In particular, $\text{PM}_{2.5}^{*}$ gains prominence in $\text{PC}_{7.2}$
during summers, reaching weights of $-0.8$. $\text{PC}_{7.3}$ largely
depends on $\text{PM}_{2.5}^{*}$ and $\text{PC}_{7.3}$ - on $\text{O}_{3}^{*}$.
$\text{PC}_{7.5}$ is overwhelmingly dependent on values of $\text{CO}^{*}$
with mean of absolute coefficients (MAC) of 0.77. $\text{O}_{3}^{*}$
and $\text{NO}_{2}^{*}$ appear to weigh in seasonally in winters
and summers respectively. Other variables appear to bring noise to
the components.

In the afternoon (right figure) we note that the decomposition of
$\text{PC}_{\text{2pm}.1}$ is more distorted. $\text{CO}^{*}$ and
$\text{NO}_{2}^{*}$ are still significant (and positively) contributors,
but their weights are now more variable (more rugged curve). Also,
$\text{O}_{3}^{*}$ is now a major contributor to $\text{PC}_{\text{2pm}.2}$,
while appears as noise in $\text{PC}_{\text{2pm}.1}$ . $\text{SO}_{2}^{*}$
is a second major contributor to $\text{PC}_{\text{2pm}.2}$. However,
its MAC dropped to 0.52 from 0.57. $\text{PM}_{2.5}^{*}$ dominates
$\text{PC}_{\text{2pm}.3}$ and $\text{PC}_{\text{2pm}.5}$. The shapes
of the remaining loading coefficients in other components are less
discernible.

When evaluated at complementary hours (figures not shown), other dynamic
loadings show similar trend in characteristics. That is higher $\overline{\text{EV}}_{1h}$
(peaking at 7am) correspond to greater linearity among loading, and
vise versa.

\begin{figure}[H]
\begin{centering}
\includegraphics[width=1\textwidth]{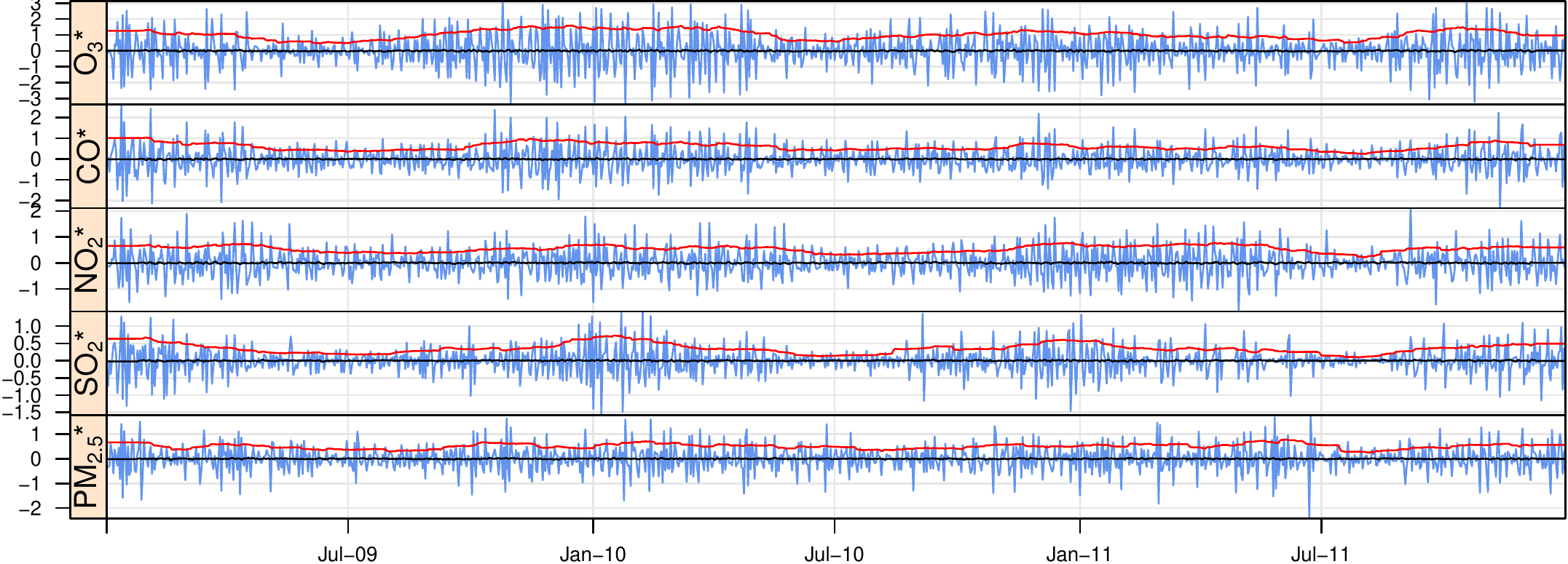}
\par\end{centering}
\caption{\label{fig:apx_NSAO} $\text{NSAO}_{\text{7am}}$. Measurements are
adjusted for DST. Overlaid curves are 45-day rolling statistics: simple
mean (black), standard deviation (red). Asterisk in $\text{O}_{3}^{*}$
is the notation for the transformed $\text{O}_{3}$ concentrations.}
\end{figure}

Loading weights control variables' participation in the make up of
the PCs. Hence, a greater (in absolute terms) loading coefficient
of a variable implies greater contribution (from the associated variable)
to the variance of the corresponding PC. So, when $v_{7..1}$ (see
Figure \ref{fig:PC Loadings}) is juxtaposed with the corresponding
$\text{EV}_{7.1}$ (see Figure \ref{fig:CEV-2D-NSAO}), we notice
the seasonal variability of $\text{NSAO}_{7}$ (see Figure \ref{fig:apx_NSAO}
in Section \ref{sec:Supplemental-material}) passing through the stable
coefficients of $v_{7..1}$ yielding a seasonal variability of $\text{PC}_{7.1}$
and $\text{EV}_{7.1}$ . While we observe this in morning hours (near
7am, when MAC peaks), this relationship is weaker in the afternoon,
especially 2pm.

Left panel of Figure \ref{fig:PC Loadings} presents $5\times5$ loadings
matrix, $\left[v_{7.pk}\in\mathbb{R}^{\mathfrak{d}}\right]_{p,k=1..5}$,
computed from a PCA on a (standardized) 45-day window sliding in time
along $\text{NSAO}_{\text{7am}}$. $\mathfrak{d}=1095$ is number
of days. Matrix columns, $\left[v_{7.pk}\right]_{\forall p}$, are
dynamic PC loadings. Matrix elements, $\left[v_{7.pk}\right]$, are
a daily TS of $k$th PC loading's coefficients (or weights), placed
in plot panels. Linear combination (at a corresponding $\text{hour}\times\text{day}$
index) of $k$th PC loading and transformed observations results in
a $k$th PC. For example, $\text{PC}_{7.1}$ is a linear combination
of weighted (transformed) pollutant concentrations, i.e. $\text{PC}_{7.1}=\sum_{p}v_{7.p1}\odot\text{NSAO}_{7.p}\in\mathbb{R}^{\mathfrak{d}}$,
where $\odot$ is a Hadamard product, and $v_{7.11}$ is a top left
(daily TS in blue) element of loading matrix and so on. Refer to \eqref{eq:PCA_Matrix_Form}
for more info. Legend values (in gray on each panel) indicate the
mean of absolute coefficients (MAC), i.e. $\overline{v}_{711}=\frac{1}{\mathfrak{d}}\sum_{d}\left|v_{7d11}\right|=.47$.
Largest MAC, $\max_{p}\overline{v}_{hpk}$, of $k$th loading sets
the direction, i.e. sign, of all $k$th loading's elements, since
signs are \emph{arbitrarily} set by many PCA computational packages
(see \texttt{prcomp()} help manual in \textbf{R}). So, $\left(\mathring{p},k\right):=\text{argmax}_{p}\overline{v}_{hpk}$
is largest MAC's location (panel). We flip signs of pointwise coefficients
via $v_{hdpk}\cdot\text{sign}\left(v_{hd\mathring{p}k}\right)$, so
as to keep $v_{hd\mathring{p}k}>0$. Finally, we smooth coefficient
series with a 45-day mean. Reflection and smoothing ease their visualization
and interpretation. Horizontal units are days in a ``mm/yy'' format
with vertical grid bars placed at 6 month increments. 

When also tried varimax orthogonal rotations of loadings, but rotated
coefficients were not materially more revealing.

\subsection{\label{subsec:Comparison-to-previous-work}Comparison to previous
work}

While application of PCA has recently gained traction in the perusal
of environmental (and meteorological) data, unfortunately, most applications
are still constrained to the static assessment. It is perspicuous
that a time-invariant PCA is unable to seize the aforementioned two-dimensional
ramifications of DPCA on a cyclostationary data, exampled with air
pollution concentration series. Static PCA assumes that an observed
sample is randomized, time ordering is unimportant and the underlying
data patterns remain constant in time \cite{ku_disturbance_1995}.

Some of the widely cited works of Statheropoulos and Abdul-Wahab (\cite{m._statheropoulos_principal_1998},
\cite{abdul-wahab_principal_2005}, respectively) rely on employment
of static PCA to dynamic air pollution data. Interestingly, the former
effort includes plots exhibiting non-stationary (seasonal) dynamics
of daily time series of raw pollutant concentrations (and meteorological
observations) and the latter discusses the diurnal dynamics of the
pollutants. Both papers (and many other efforts) stepped in the direction
of dynamic analysis by applying PCA separately to winter and summer
seasons (Statheropoulos) and day and night time (Abdul-Wahab). Still
this assumes that the data structure wobbles between two constant
states, which is not the case with environmental and meteorological
data. Moreover, there is limited discussion of PCA assumptions and
robustness of the results. The latter paper utilizes standardized
$\log\left(\text{ozone}\right)$ observations, but appears to leave
other variables intact. The former publication does not mention any
transformation of the notably cyclical observation series (see figures
therein). Not surprisingly the $\text{EV}_{k}^{\text{static}}$ in
both works remain low, under 35\% for $\text{EV}_{1}$.

We consider our work an improved and proper extension of these two
papers in application of PCA. In fact, when we employed their methods
to our normalized set (with winter/summer and day/night observations
identified analogously), we discovered a greatly improved $\text{EV}_{k}^{\text{static}}$,
as shown below in Figure \ref{fig:Prior_work}. Seasonal cycles appear
much stronger in our work (see Figure \ref{fig:CEV-2D-NSAO}) and
summer/winter $\text{EV}_{k}^{\text{static}}$ appear to capture this
with similar pattern strength in winter observations. Decomposition
of day and night observations is less informative, likely due to the
hours chosen by the authors (6am-5pm as day and remainder as night).
Our analysis reveals the diurnal (local) dynamics among variables
and suggests clustering night and morning hours separately from afternoon
hours. In fact, it may be helpful to have three groups: night, morning
and afternoon. Naturally, such discovery may go unnoticed without
performing our DPCA technique on each hour of the day. 

\begin{figure}[H]
\begin{centering}
\begin{tabular}{r|rrrr}
 & $\text{EV}_{1}$ & $\text{CEV}_{2}$ & $\text{CEV}_{3}$ & $\text{CEV}_{4}$\tabularnewline
\hline 
summer & .50 & .70 & .84 & .96\tabularnewline
winter & .58 & .78 & .88 & .96\tabularnewline
\hline 
daytime & .47 & .69 & .86 & .96\tabularnewline
night time & .46 & .68 & .86 & .97\tabularnewline
\end{tabular}
\par\end{centering}
\caption{\label{fig:Prior_work}Cumulative explained variance based on static
PCA work of Statheropoulos (lagging $\text{O}_{3}$ observations;
summer vs winter) and Abdul-Wahab (contemporaneous analysis; night
vs day). We analogously aggregated hourly $\text{SAO}$ data; then
normalized with log differencing defined in \eqref{eq:Diff-Log} and
employed PCA.}
\end{figure}

Finally, dynamic PCA yields a greater information, when compared to
static PCA, about seasonal patterns in the variables, with $\text{EV}_{h.1}$
reaching $70-75\%$ (see Figure \ref{fig:CEV-2D-NSAO}) in winter
nights of our dataset. Our DPCA application enables a higher quality
air pollution analysis targeted at a particular season or time of
day. The components can further be used in regression or other statistical
methods for the purposes of quality prediction and air pollution studies.

\section{Conclusion}

The objective of our study was to highlight the dynamic nature of
air pollutants. We accomplished this objective by applying non-local
DPCA at each of the 24 hours of a day to investigate Houston's air
pollution profile. Thus, we constructed a two dimensional analysis
over $\text{hours}\times\text{days}$ domain, essentially separating
diurnal and seasonal cycles. We have discovered that daylight savings
have an insignificant impact on the analysis. We then chose and tested
a suitable normalizer (log differencing) that transforms our data
set $\text{SAO}_{h}$ to an approximately multivariate normal, $\text{NSAO}_{h}$
(percent change in averaged pollutant concentrations). Still, we briefly
compared (at the aggregate level of MAC) DPCA done on $\text{NSAO}_{h}$
versus those on the original $\text{SAO}_{h}$ and (frequently used)
$\text{LSAO}_{h}$ datasets. We presented the dynamic explained variance
and loadings at each hour. 

The key finding was that the air pollution profile remains non-constant
throughout a day and throughout a year. The best EV is achieved in
the morning (around 7am), when loading coefficients exhibit linear
and consistent (non-local) structure regardless of the season. $\text{PC}_{h.1}$
captures seasonal profile at any hour $h$, although its seasonal
structure is poorest in the afternoon. This is when many of the dynamic
loadings are least meaningful as well. 

The novelty of this paper is a new and proper application of PCA to
an air pollution dataset. We show that given the nature of complex
pollutant associations with daily and annual cycles, it's not only
important, but also highly worthwhile to apply PCA on a subset of
cyclostationary data. Such practice identifies patterns of strengthening
and weakening of correlations among studied variables throughout a
day or a year.

We then compared our results to existing (static PCA) research efforts
and concluded that DPCA unveils a much richer and more complete dynamics
of the analyzed data.

This work does not attempt to build predictors, reduce dimensionality,
or construct air pollution indicators. Yet, the determined uncorrelated
PCs are suitable for application of further extensions such as regression,
self organizing maps (SOM), artificial neural networks (ANN), and
other techniques. 

\section*{References}

\bibliographystyle{plain}
\addcontentsline{toc}{section}{\refname}\bibliography{DPCA}

\section{Appendix}

\subsection{Daylight saving time (DST)}

Most businesses operate in local time, setting pace for traffic hours
and, hence, pollutant emissions \cite{gurevitz_daylight_2005,yacker_daylight_1998,hecq_daylight_1993}.
Likewise, most of the Americas use DST to extend evening hours into
daylight at the expense of morning hours. In particular, Houston,
and the whole of Texas, are in the Central Time (CT) zone. This zone
follows the Central Daylight Time (CDT) convention from a ``jump''
day in mid-March to a ``compression'' day in early November and
the Central Standard Time (CST) convention for the remainder of the
calendar year. CDT and CST are 5 and 6 hours (respectively) behind
Coordinated Universal Time (UTC), which is  Greenwich Mean Time (GMT),
which does not observe DST.

Initially, our raw data is indexed with UTC-6:00 (i.e. ignores CST/CDT
adjustments) uninterrupted (no jumps or compressions) hourly increments.
However, the relation of pollutants to traffic and diurnal human activity
prompts the investigation of the effect of DST \cite{munoz_morning_2007}
on PCA outcome. Apparently, the use of the CST/CDT index has made
only a diminutive amelioration (of 0.01\%) in $\text{EV}_{1}$. The
whole improvement came from the PCA of a moving window over the jump
and compression days.

Still we carry on the analysis in local (i.e. CST/DST) time zone.
This results in one missing 2am observation when CDT goes into effect
on jump day, and one duplicate when CST takes effect on compression
day in each year. For simplicity, we interpolate the former and delete
the later. 

\begin{figure}[h]
\begin{centering}
\includegraphics[width=0.49\columnwidth]{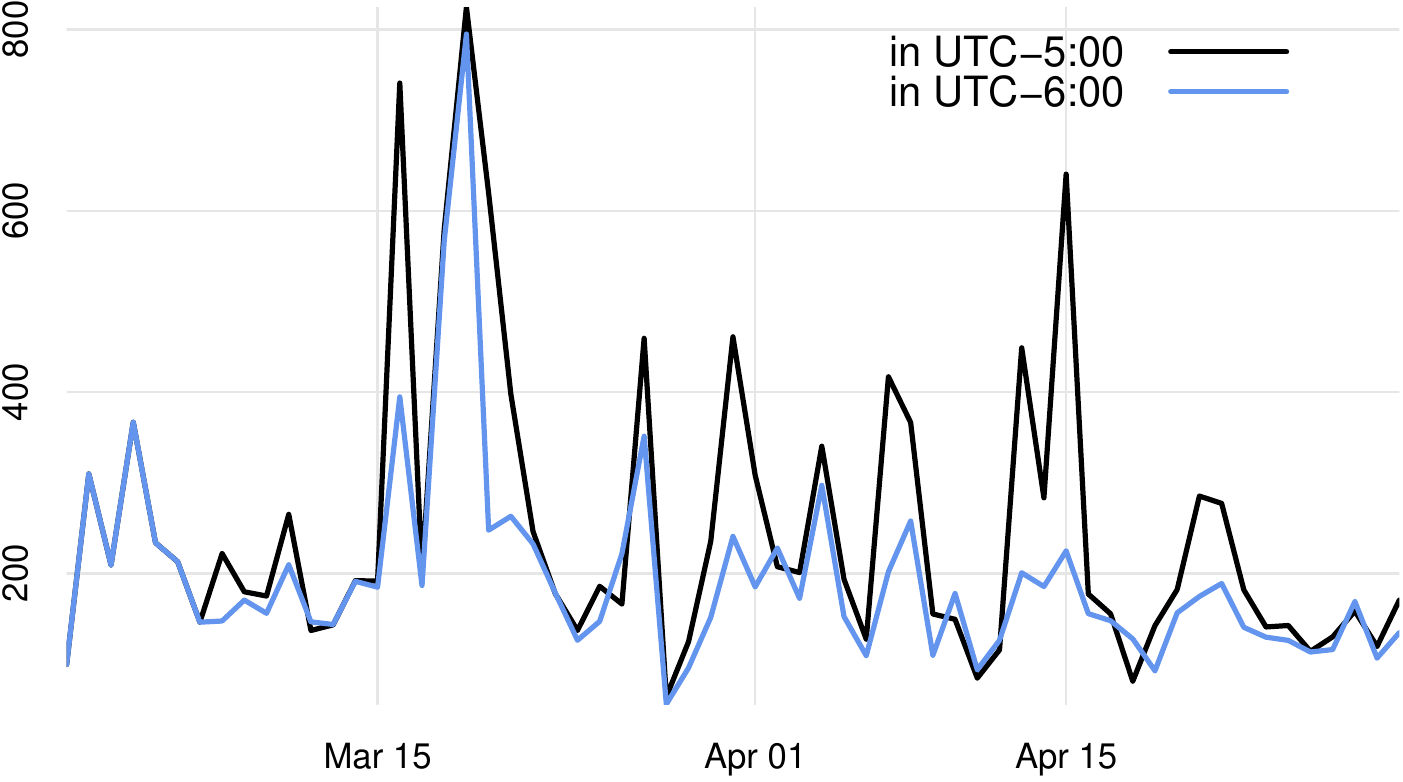}
\hfill{}\includegraphics[width=0.49\columnwidth]{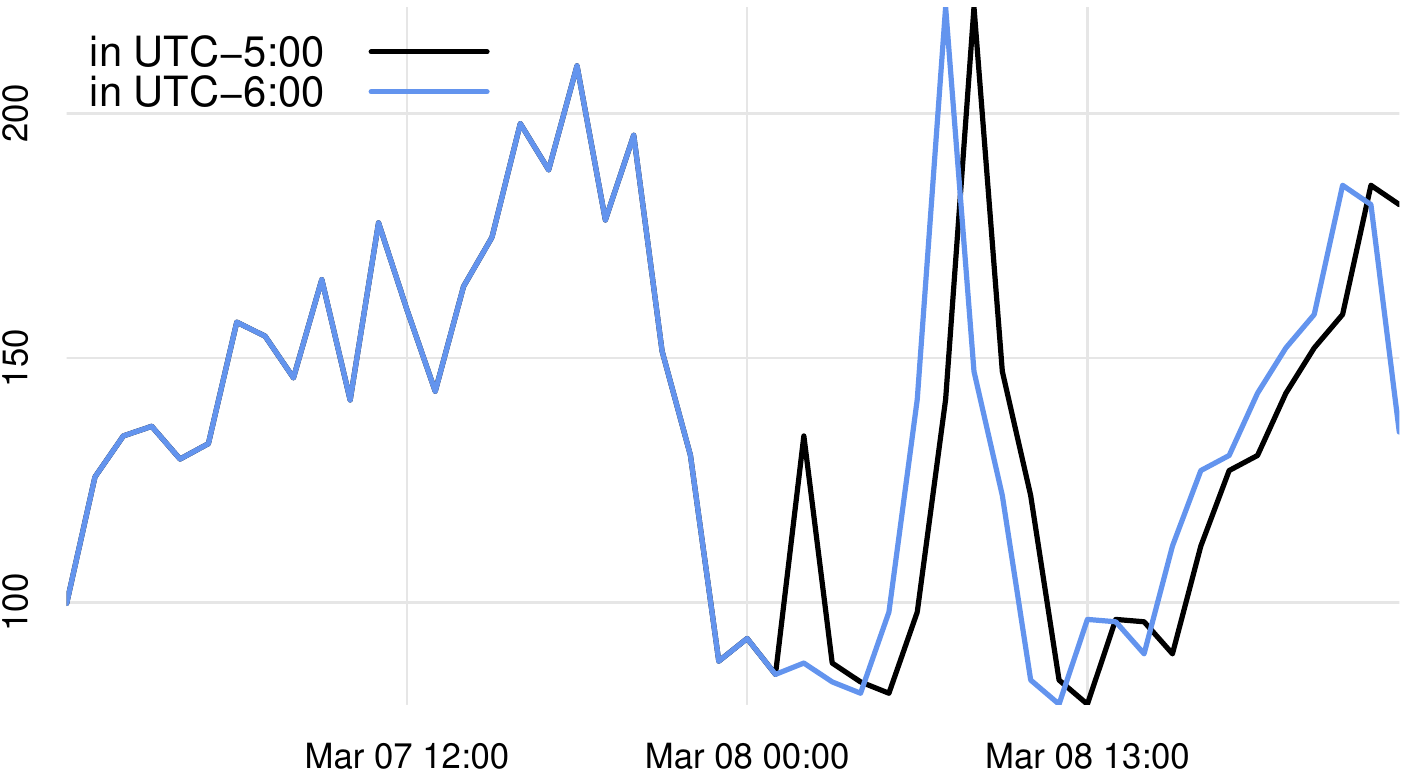}
\par\end{centering}

\caption{\label{fig:CST-CDT} DST effect on CO measurements (in ppb). In 2009
CST/CDT jump occurred on March 8 at 2am. On jump day local time shifts
forward by one hour from $\text{UTC-6:00}$ to $\text{UTC-5:00}$,
i.e. $\text{1am CST}\to\text{2am CDT}$ and so on. \emph{Left}: Daily
measurements at 8am. \emph{Right}: Hourly measurements around time
change (jump event).}
\end{figure}

Figure \ref{fig:CST-CDT} exemplifies a jump in observations when
time shifts from CDT to CST. The left panel shows \emph{non-local}
observations, i.e. daily concentrations at a fixed time (at a 24 hour
lag). The right panel shows \emph{local} observations, i.e. consecutive
hourly concentrations, as defined in \cite{davis_outlier_2006}. Note
that (averaged) non-local CO levels remain higher for the adjusted
data at 8am, i.e. black curve is atop blue curve on the left panel.
This is expected, since the CST/CDT-indexed concentrations reflect
morning traffic's CO emissions faster than the UTC-6:00 indexed measurements.
The right panel shows a shadow effect as unadjusted concentrations
remain one hour behind the adjusted ones.

\newpage{}

\section*{\label{sec:Supplemental-material}Supplemental material}

\begin{figure}[H]
\begin{centering}
\includegraphics[width=1\textwidth]{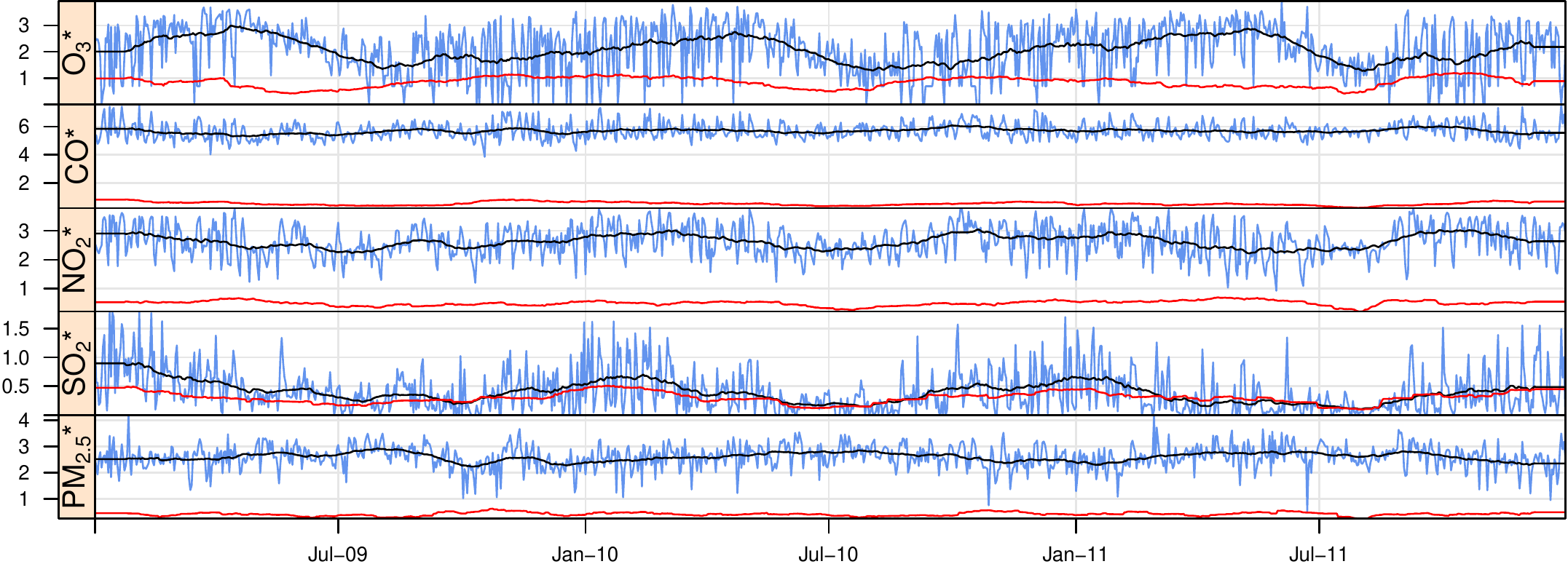}
\par\end{centering}
\caption{\label{fig:apx_LSAO} $\text{LSAO}_{\text{7am}}$. Measurements are
adjusted for DST. Overlaid curves are 45-day rolling statistics: simple
mean (black), standard deviation (red). Asterisk in $\text{O}_{3}^{*}$
is the notation for the transformed $\text{O}_{3}$ concentrations.}
\end{figure}

\begin{figure}[H]
\includegraphics[width=0.327\textwidth]{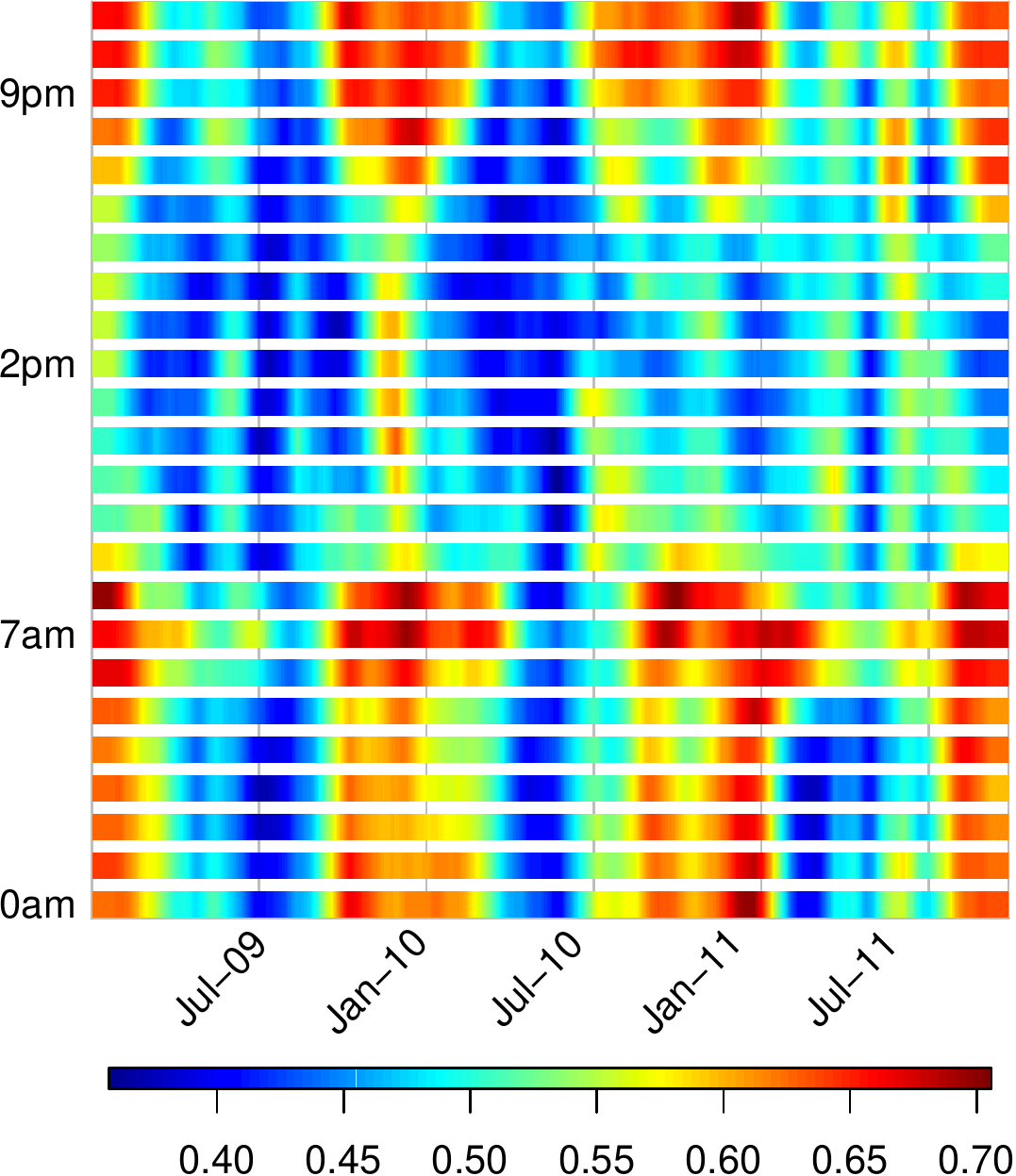}\hfill{}\includegraphics[width=0.327\textwidth]{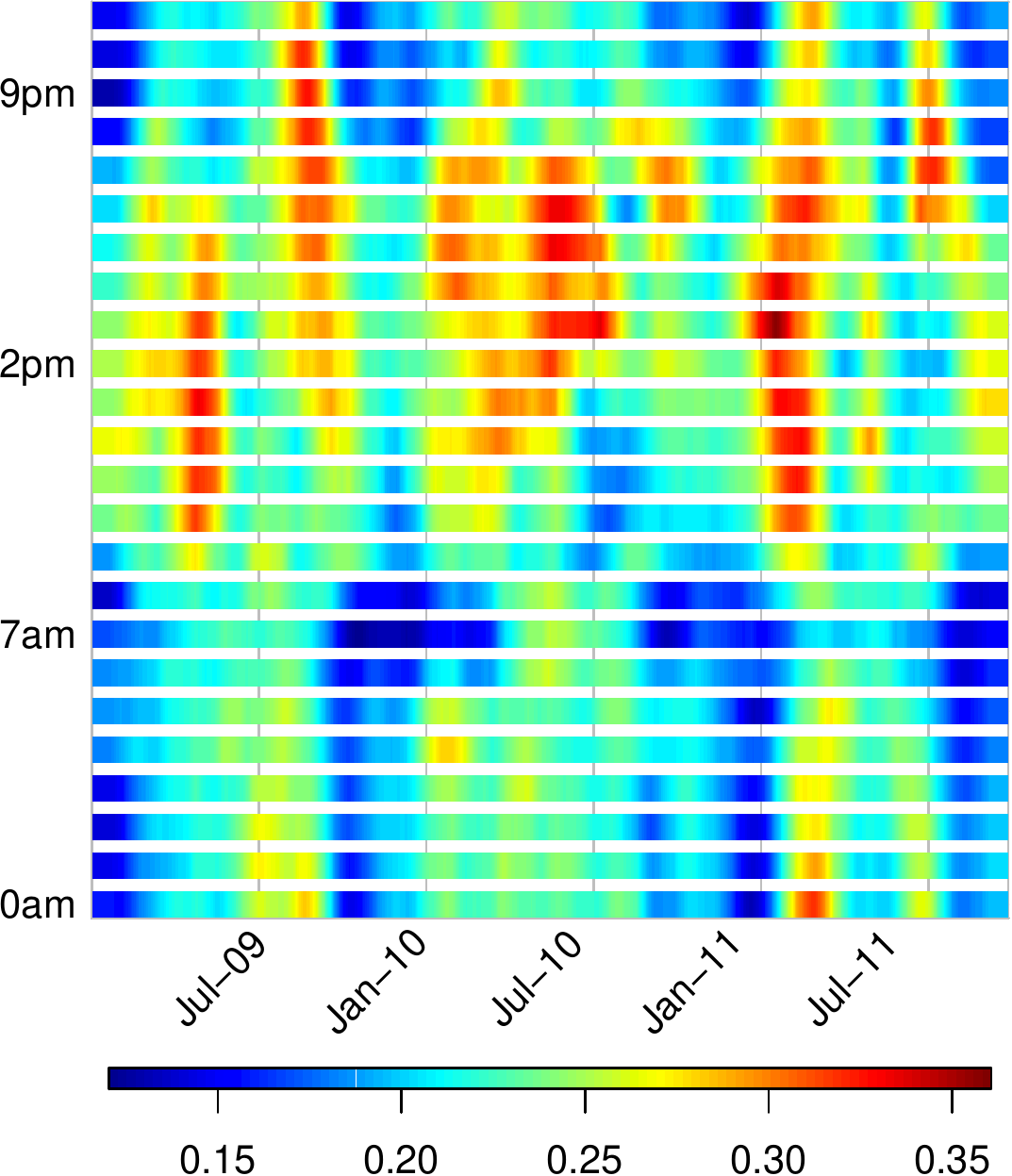}\hfill{}\includegraphics[width=0.327\textwidth]{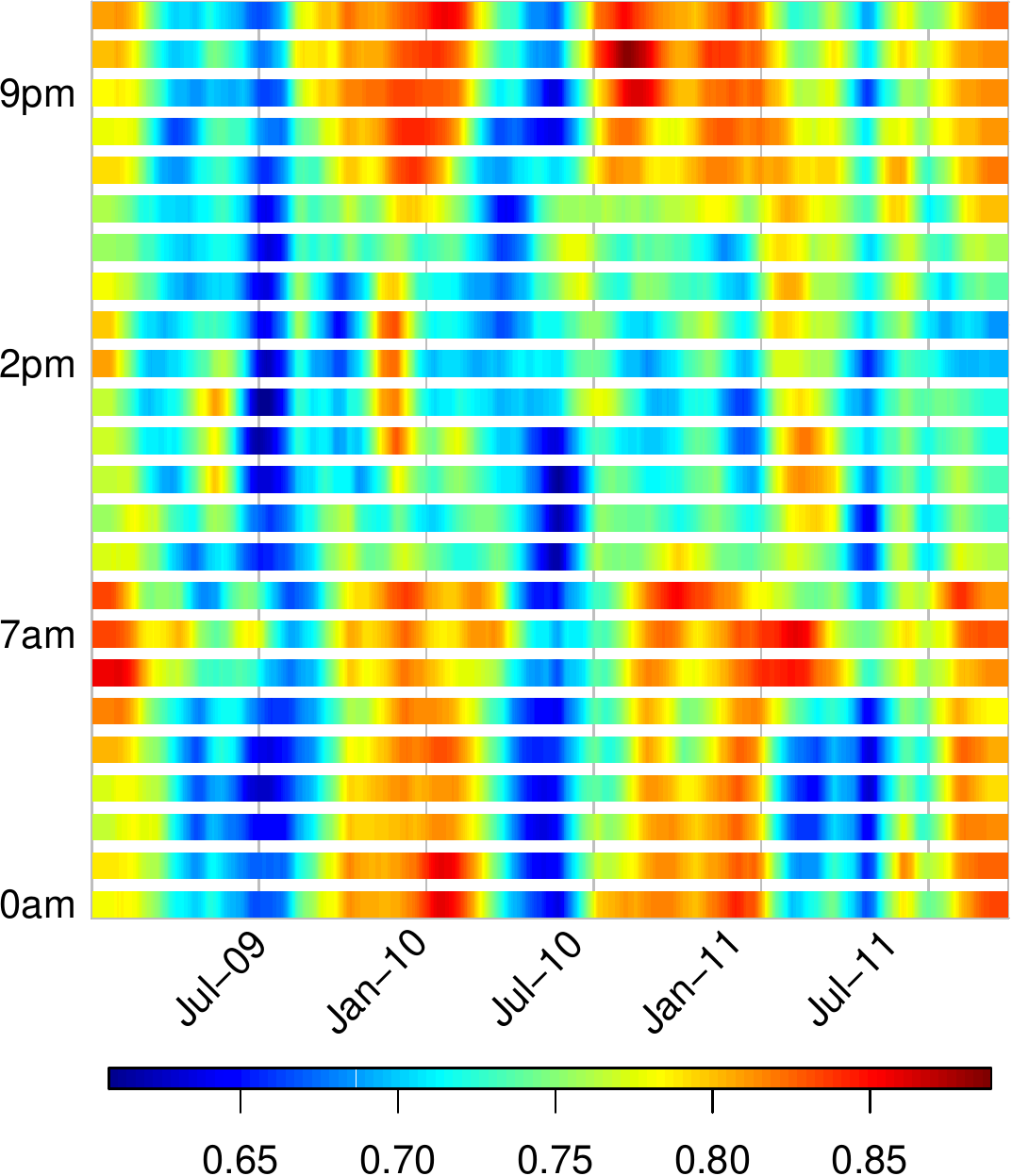}

\caption{\label{fig:apx_EV-2D-SAO} Heatmap of dynamic explained variance (EV)
from $\text{PC}_{h.1}$ and $\text{PC}_{h.2}$ of $\text{SAO}_{h}$
and $\text{SAO}_{h}$, $h=0..23$. \emph{Left} and \emph{center}:
non-cumulative for the first two components. \emph{Right}: cumulative
for both components.}
\end{figure}

\begin{figure}[H]
\includegraphics[width=0.327\textwidth]{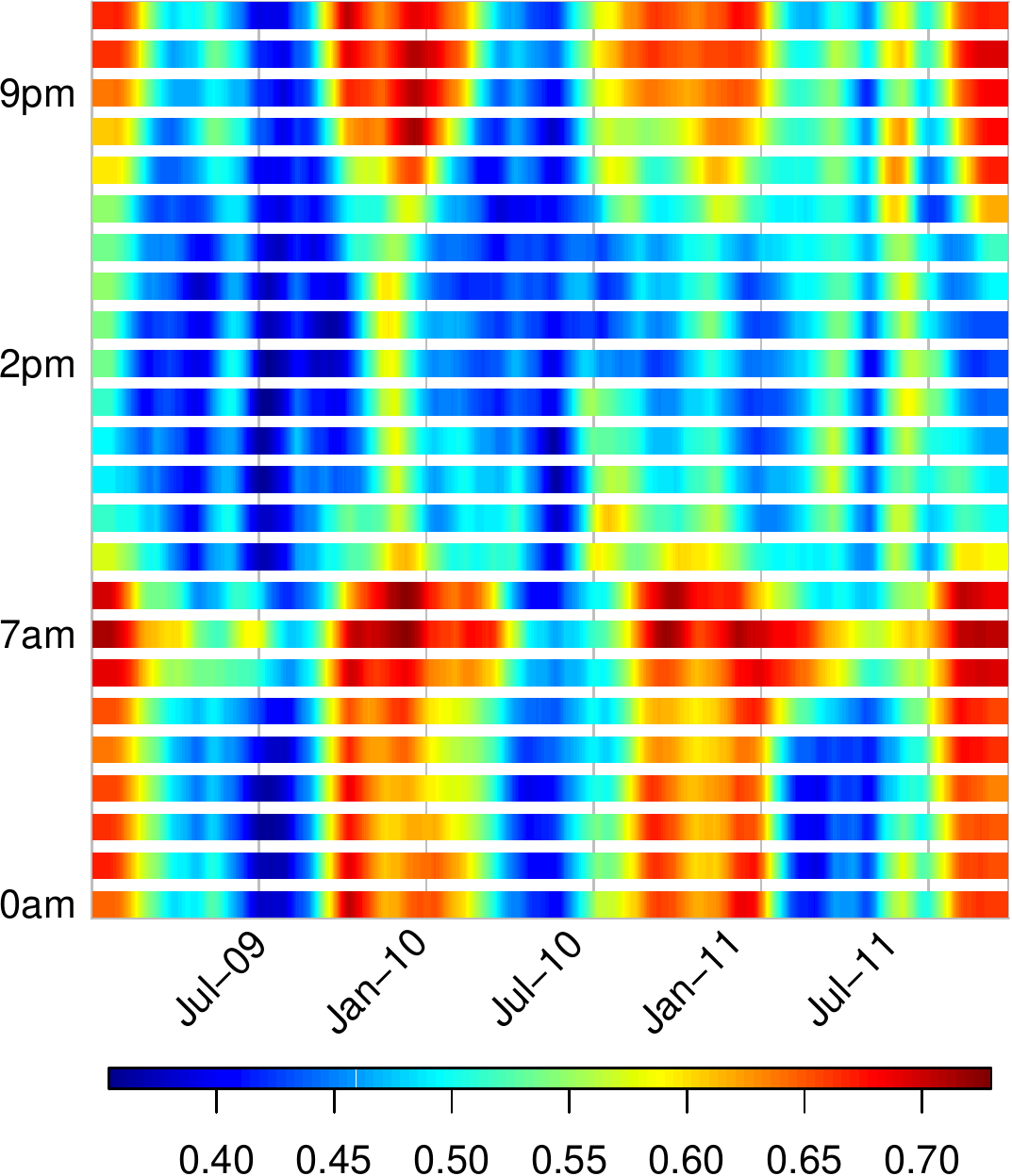}\hfill{}\includegraphics[width=0.327\textwidth]{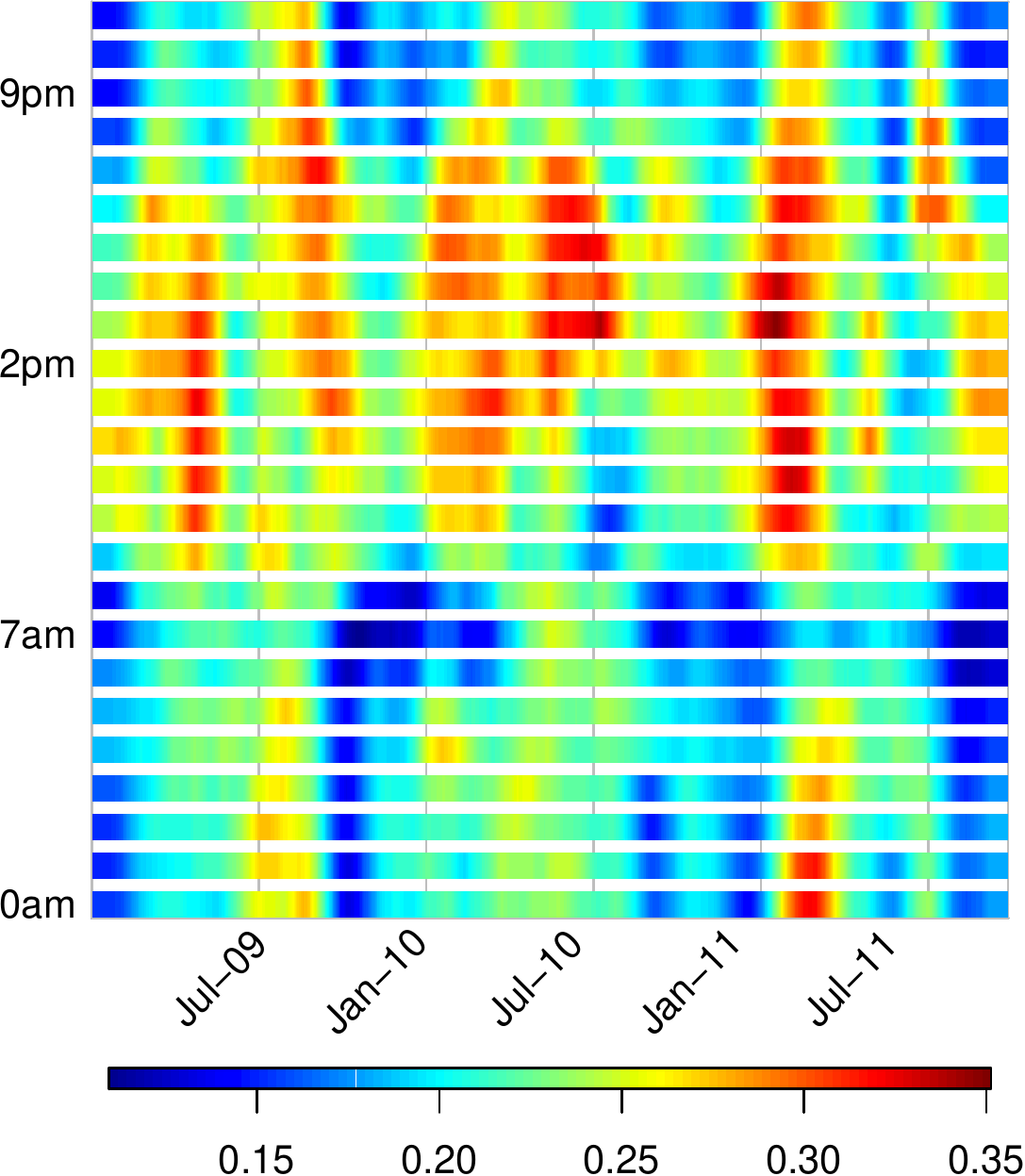}\hfill{}\includegraphics[width=0.327\textwidth]{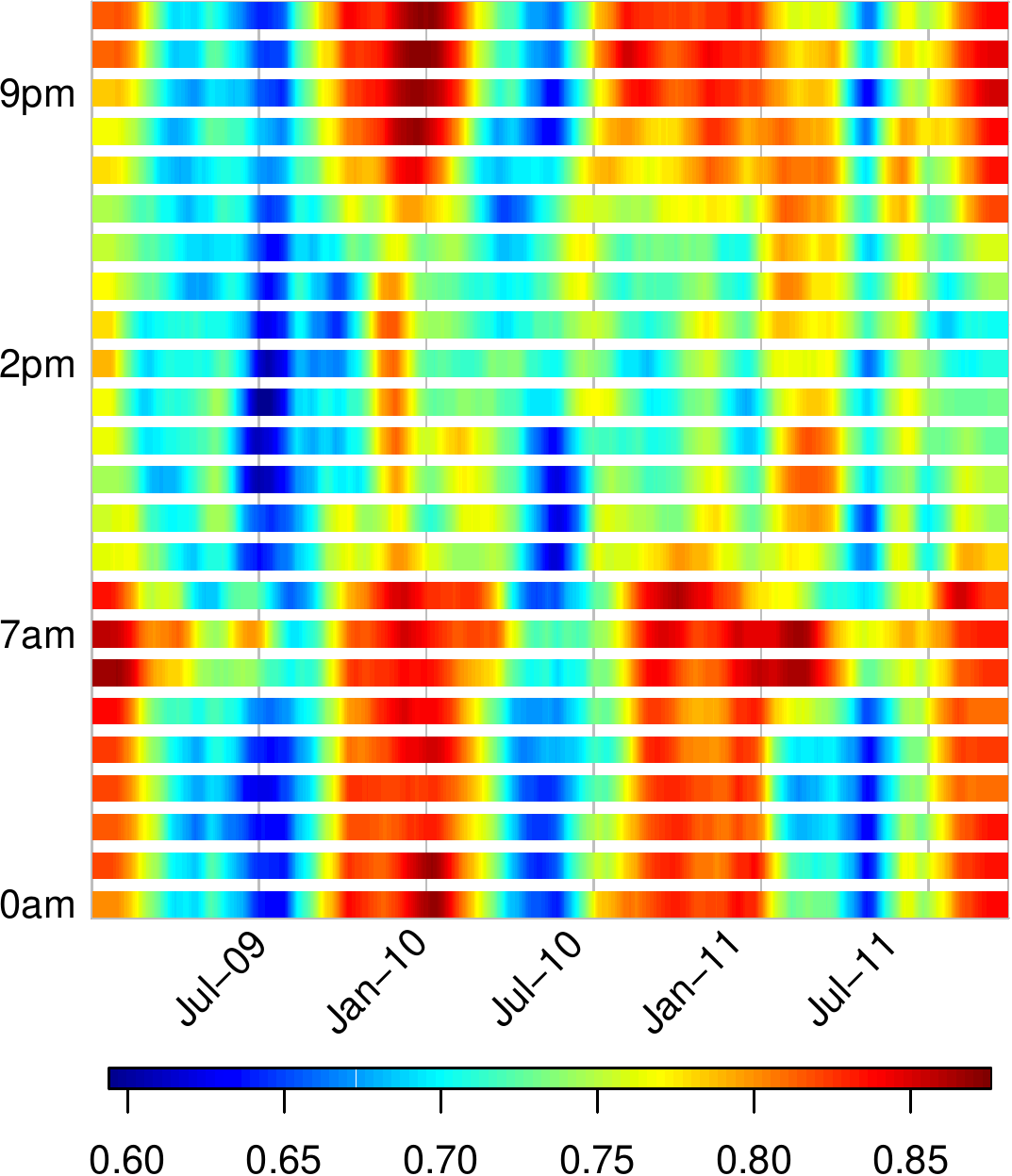}

\caption{\label{fig:apx_EV-2D-LSAO} Heatmap of dynamic explained variance
(EV) from $\text{PC}_{h.1}$ and $\text{PC}_{h.2}$ of $\text{SAO}_{h}$
and $\text{LSAO}_{h}$, $h=0..23$. \emph{Left} and \emph{center}:
non-cumulative for the first two components. \emph{Right}: cumulative
for both components.}
\end{figure}

Dotted lines (in matching colors) represent the non-local means across
the whole 3 year period. Vertical units are proportions on 0-1 scale
(1 is 100\% contribution to variance).

\begin{figure}[H]
\begin{centering}
\includegraphics[width=1\textwidth]{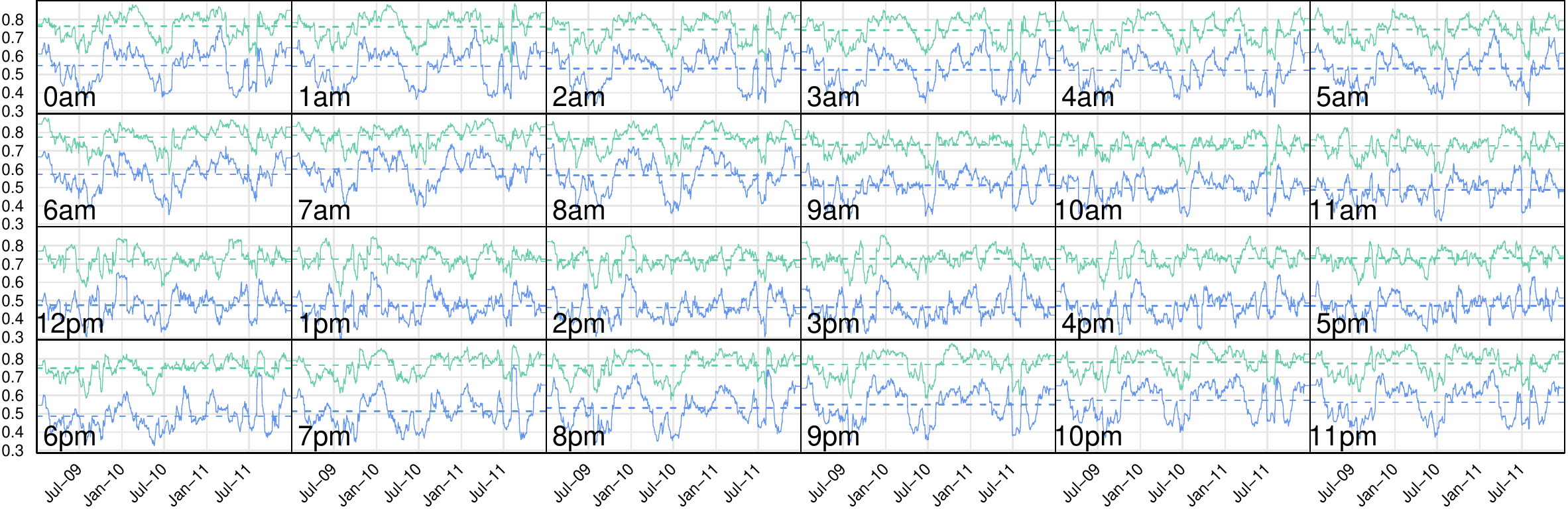}
\par\end{centering}
\caption{\label{fig:apx_CEV24-SAO}CEV from $\text{PC}_{1}$ and $\text{PC}_{2}$
of $\text{SAO}_{h},h=0..23$ hours. }
\end{figure}

\begin{figure}[H]
\begin{centering}
\includegraphics[width=1\textwidth]{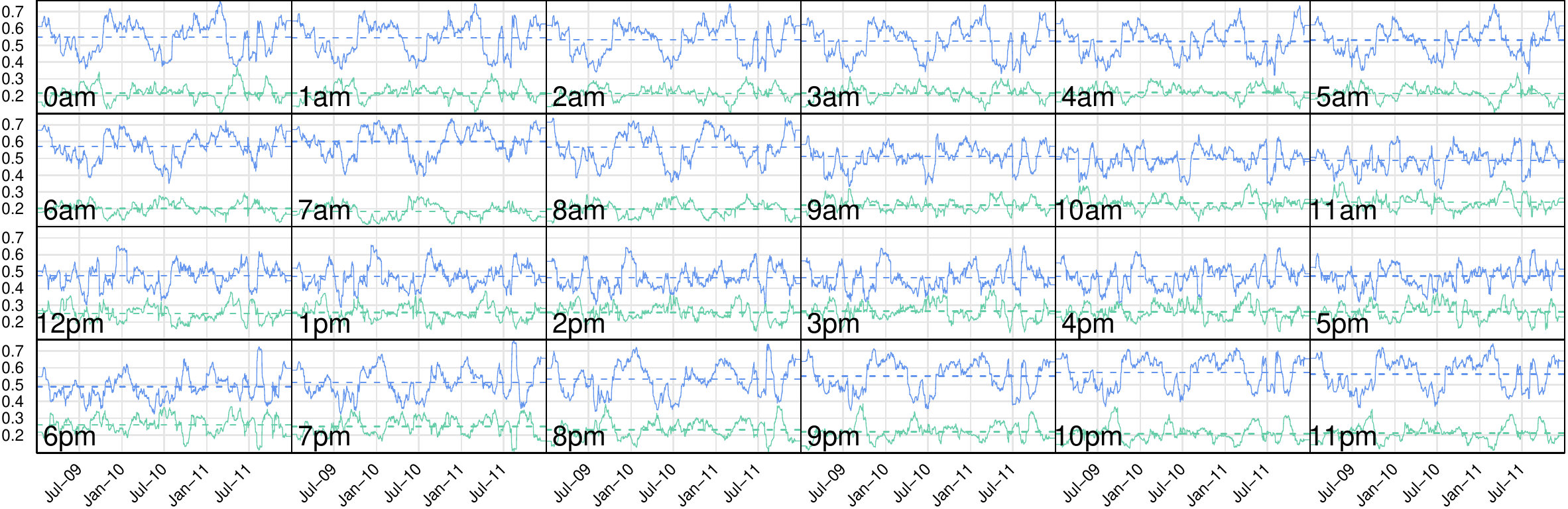}
\par\end{centering}
\caption{\label{fig:apx_EV24-SAO}EV from $\text{PC}_{1}$ and $\text{PC}_{2}$
of $\text{SAO}_{h},h=0..23$ hours. }
\end{figure}

\begin{figure}[H]
\begin{centering}
\includegraphics[width=1\textwidth]{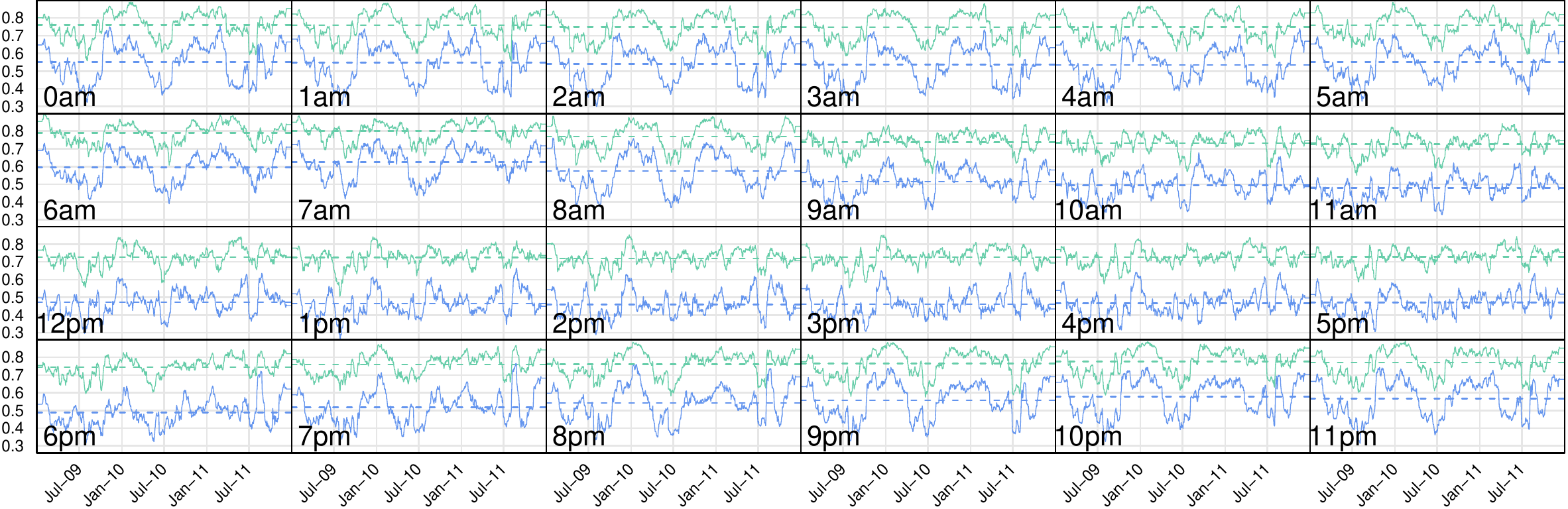}
\par\end{centering}
\caption{\label{fig:apx_CEV24_LSAO}CEV for the first two components ($\text{EV}_{ihd}$,
$i=1,2$) of $\text{LSAO}_{h},h=0..23$ hours.}
\end{figure}

\begin{figure}[H]
\begin{centering}
\includegraphics[width=1\textwidth]{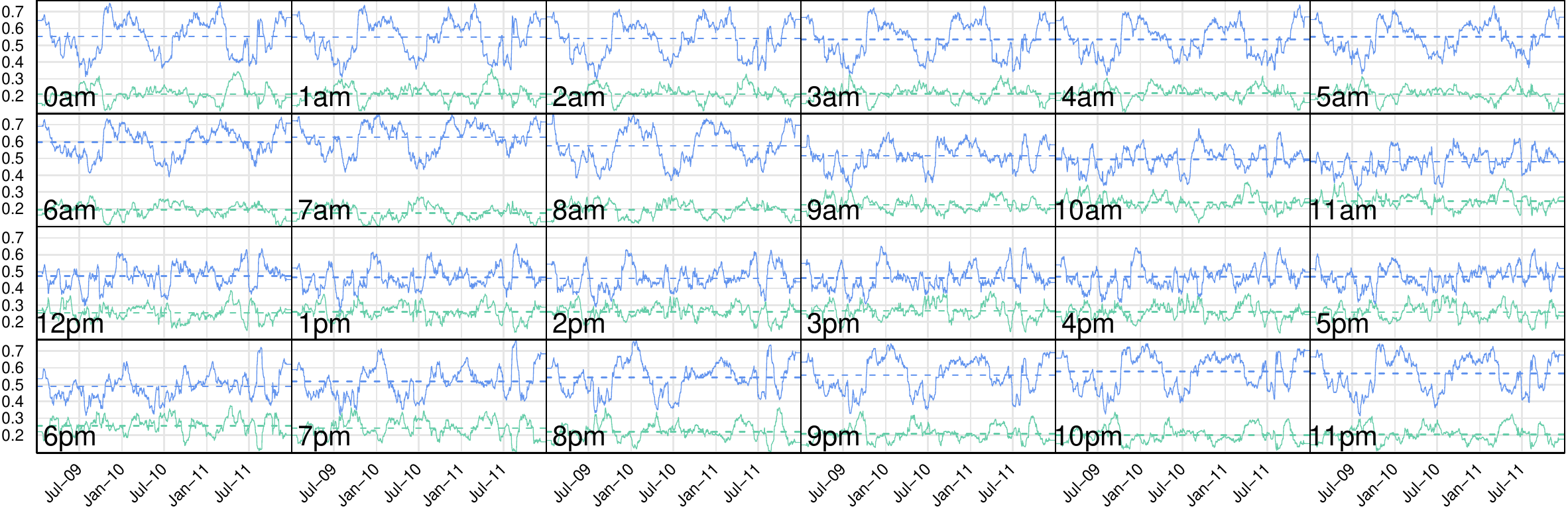}
\par\end{centering}
\caption{\label{fig:apx_EV24_LSAO} EV for the first two components ($\text{EV}_{ihd}$,
$i=1,2$) of $\text{LSAO}_{h},h=0..23$ hours.}
\end{figure}

\begin{figure}[H]
\begin{centering}
\includegraphics[width=1\textwidth]{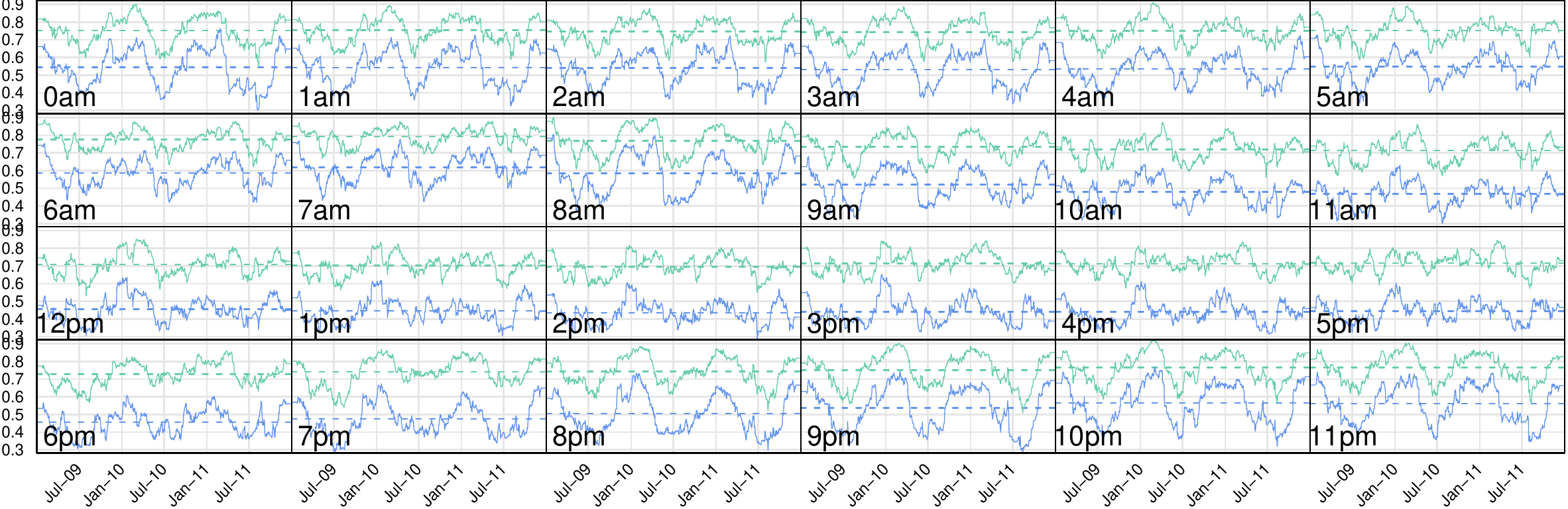}
\par\end{centering}
\caption{\label{fig:apx_CEV_NSAO} CEV for the first two components ($\text{EV}_{ihd}$,
$i=1,2$) of $\text{NSAO}_{h},h=0..23$ hours.}
\end{figure}

\begin{figure}[H]
\begin{centering}
\includegraphics[width=1\textwidth]{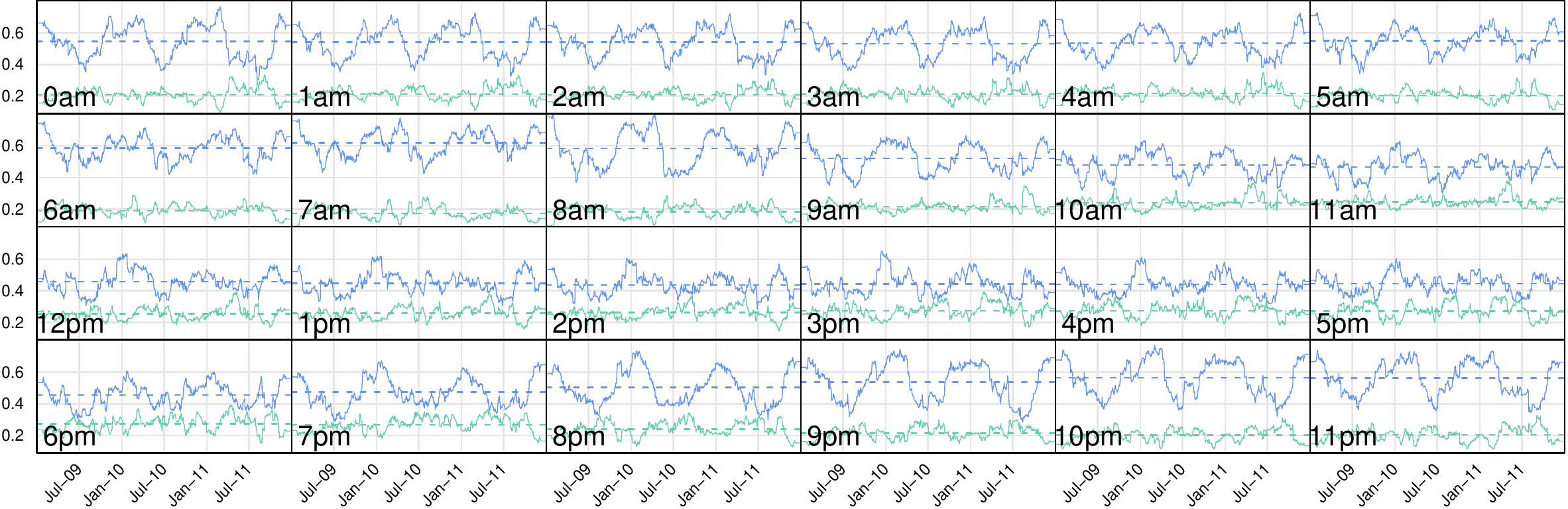}
\par\end{centering}
\caption{\label{fig:apx_EV_SAO} EV for the first two components ($\text{EV}_{ihd}$,
$i=1,2$) of $\text{NSAO}_{h},h=0..23$ hours.}
\end{figure}

\end{document}